\newif\ifconfver
\confvertrue
\ifconfver
\documentclass[10pt,twocolumn,twoside]{IEEEtran}
\else
\documentclass[11pt,draftcls,onecolumn]{IEEEtran}
\fi
\usepackage{hyperref}
\usepackage{makecell}
\usepackage{bbding}
\usepackage{color,graphicx}
\usepackage{float}
\usepackage{multirow,amsmath,epsfig,amsfonts,amssymb,psfig,graphics,psfrag,theorem,calc,url,bm,cite}
\usepackage{stfloats,hyperref}
\usepackage{algorithm,algorithmic}
\usepackage{diagbox}
\usepackage{hhline}
\usepackage{subfigure}
\usepackage{pifont}
\usepackage{physics}
\usepackage{booktabs}
\usepackage{tikz}
\usepackage{multirow}
\usetikzlibrary{quantikz}
\usepackage{arydshln}
\usepackage{orcidlink}
\usepackage{mathtools}

\makeatletter
\def\multilimits@{\bgroup
	\Let@
	\restore@math@cr
	\default@tag
	\baselineskip\fontdimen10 \scriptfont\tw@
	\advance\baselineskip\fontdimen12 \scriptfont\tw@
	\lineskip\thr@@\fontdimen8 \scriptfont\thr@@
	\lineskiplimit\lineskip
	\vbox\bgroup\ialign\bgroup\hfil$\m@th\scriptstyle{##}$\hfil\crcr}
\def\Sb{_\multilimits@}
\def\endSb{\crcr\egroup\egroup\egroup}
\makeatother

	{\end{list}}

\newlength{\twidth}
\ifconfver
\else
\fi

\definecolor{orange}{RGB}{255,107,0}

\theorembodyfont{\rmfamily}

{\begin{list}{}{
			\settowidth{\labelwidth}{\mbox{\textnormal{#1}}}%
			\setlength{\leftmargin}{\labelwidth+\labelsep}}}%
	{\end{list}}

\definecolor{orange}{RGB}{255,107,0}

\ifconfver

\author{Lino Garda Denaro \orcidlink{0000-0001-5367-5392}, and
Chia-Hsiang Lin \orcidlink{0000-0002-4865-2329},~\IEEEmembership{Senior Member,~IEEE}}

\title{Monotone Piecewise-Linear Harmonization of Multitemporal Multispectral Surface Reflectance With Unsupervised Deep Multiview Pseudoinvariant Features Learning

\thanks{This study was supported by the Emerging Young Scholar Program (namely, the 2030 Cross-Generation Young Scholars Program) of National Science and Technology Council (NSTC), Taiwan, under Grant NSTC 115-2628-E-006-001.
We thank the National Center for Theoretical Sciences (NCTS) and the National Center for High-performance Computing (NCHC) for providing the computing resources.
\textit{(Corresponding Author: Chia-Hsiang Lin)}}
\thanks{Lino Garda Denaro is with the Research Center for Geoinformatics, National Research and Innovation Agency (BRIN), Cibinong, Indonesia and with the Department of Electrical Engineering, National Cheng Kung University (NCKU), Tainan, Taiwan.
(e-mail: dlinogarda@brin.go.id \& dlinogarda@gmail.com).}
\thanks{C.-H. Lin is with the Department of Electrical Engineering, National Cheng Kung University (NCKU), Tainan 70101, Taiwan (R.O.C.)
(e-mail: chiahsiang.steven.lin@gmail.com).}
}
\fi

\begin{document}

	\bibliographystyle{IEEEtran}
	\maketitle
	\ifconfver \else \vspace{-0.5cm}\fi

\begin{abstract}
Surface reflectance products derived from optical satellite imagery may exhibit systematic inconsistencies across acquisition dates, because atmospheric correction is generally performed independently for each image. Such residual biases distort the temporal behavior of invariant ground targets and reduce the reliability of multitemporal surface reflectance analysis. This study proposes Deep Multitemporal Generalized Canonical Correlation Analysis (DeepMGCCA), an unsupervised deep multiview pseudoinvariant feature learning framework that integrates Deep Generalized Canonical Correlation Analysis (GCCA) with Multivariate Alteration Detection (MAD) to identify deep-pseudoinvariant features (DIFs) for the removal of surface reflectance biases across multitemporal satellite observations. Unlike conventional MAD-based approaches that rely on linear GCCA, the proposed framework replaces date-specific linear projections with a shared convolutional encoder. 
%
%
%
%
The framework learns deep invariant features (DIFs) directly from pixel-weighted canonical variates via iterative reweighting strategy, while a reconstruction regularizer preserves surface reflectance information and prevents degenerate collapse from suppressing genuine temporal variability.
%
%
Experiments were conducted on four annual Landsat-8/9 and Sentinel-2 acquisitions across four regions of interest (ROIs) to evaluate four DIF-based harmonization transforms.
%
%
A monotone piecewise-linear transform derived from DIFs quantiles consistently achieved the largest bias reduction, lowering the between-date spectral angle by up to {2.54}{$^\circ$} and RMSE by 0.015, and raising PSNR by up to 4.04~dB. These findings demonstrate the potential of deep invariant feature learning for multitemporal surface reflectance bias removal while highlighting the importance of preserving physically meaningful temporal and spectral information during nonlinear representation learning.

\end{abstract}


	\ifconfver \else \vspace{-0.0cm}\fi

	\ifconfver \else \vspace{-0.5cm}\fi

	\ifconfver \else  \fi

\section{Introduction}\label{sec: introduction}

Multitemporal optical remote sensing analysis, including land-cover change mapping and disturbance monitoring \cite{Bovolo2007polar,Wu2014sfa}, and the reconstruction of dense reflectance time series \cite{Gao2006blending}, requires that reflectance differences between acquisition dates reflect genuine surface changes rather than differences in illumination geometry \cite{Latifovic2003brdf}, atmospheric conditions \cite{Vermote1997sixs}, sensor calibration \cite{Markham2004landsat}, or phenological background.
Relative radiometric normalization is the conventional solution that uses a set of pseudo-invariant features (PIFs), or pixels thought to be radiometrically stable throughout the time series, are used to calibrate one date to another or to a common reference. The concept originates in \cite{SCHOTT19881, Xu2023}, which selected spectrally stable man-made targets manually, and was subsequently developed into the radiometric-control-set methodology \cite{HALL199111}. Fully automatic variants, in which the invariant set is estimated from the imagery itself rather than delineated by an analyst, were established by Liu et al., and Canty et al. \cite{CANTY2004441,Liu202000} and Canty and Nielsen \cite{CANTY20081025, NielsenAasbjerg2007}. In all of these, the reliability of the normalization is bounded by the reliability of the PIF selection step, which makes automated, statistically principled PIFs extraction a problem of independent importance.
Recent studies have attacked radiometric consistency from the cross-sensor side, deriving global transformation functions between Landsat-8 and Sentinel-2 reflectance products \cite{wang2025harmu, Xie2022global}, harmonizing the two sensors spectrally through matching-band adjustment \cite{Shen2024spectral}, harmonizing multispectral and panchromatic VHR spaceborne imagery \cite{Alparone2025harmonization}, and quantifying cross-sensor consistency of surface reflectance \cite{Zhao2026terrain-aware}, to the point that harmonized Landsat--Sentinel-2 time series have become a standard input to downstream land-cover applications \cite{Lu2025mapping}. These approaches harmonize across sensors at a given epoch, the residual interdate bias that remains inside each surface reflectance product still requires multitemporal, statistically principled harmonization, which is the problem addressed in this study.

Multivariate Alteration Detection (MAD) \cite{NIELSEN19981, DU2002123} provides a statistically grounded solution by projecting two multispectral acquisitions into a set of canonical components derived through two-view Canonical Correlation Analysis (CCA) \cite{hotelling1992relations}. These components are ordered according to their bitemporal correlation, allowing temporal differences to be represented in a transformed space that emphasizes change-related variability. Under the no-change hypothesis, the squared differences between corresponding MAD variates, normalized by their variances, approximately follow a chi-square distribution, thereby enabling pixel-level change significance to be assessed in a principled manner. The Iteratively Reweighted MAD (IR-MAD) framework \cite{Liu2015, NielsenAasbjerg2007} further improves robustness by assigning lower weights to pixels that are more likely to have changed during successive covariance re-estimation steps. This iterative procedure progressively enhances the discrimination between invariant and changed observations and provides a widely adopted basis for chi-square thresholding of the resulting change-magnitude measure. Other statistically motivated approaches for thresholding multitemporal difference images include the parameter-estimation method proposed by Bruzzone and Prieto \cite{BruzzonePrieto2000} and the PCA-based k-means strategy of Celik \cite{CelikTurgay2009}. However, unlike MAD-based methods, these approaches do not produce an ordered set of invariance-related components that can be directly exploited for pseudo-invariant feature selection.

CCA provides the foundation for MAD by identifying maximally correlated linear combinations between two sets of variables from bi-temporal images. Extending this concept to multitemporal image sequences, Nielsen \cite{Nielsen2002} introduced multiset CCA, where each acquisition date can be treated as a distinct data view. This naturally motivates the use of Generalized CCA (GCCA) to jointly model multiple image views. Early developments in multiset CCA were introduced by Horst \cite{Horst_1961, Tenenhaus_Tenenhaus_2011, lin2019, denaro2026}. Accordingly, a fully connected GCCA formulation provides a direct multitemporal generalization of the two-date MAD framework while retaining its chi-square-based change statistic and iterative reweighting strategy for identifying temporally stable observations.

All of the above formulations remain fundamentally linear, since each acquisition date is projected through a fixed matrix transformation applied independently to every pixel and therefore cannot explicitly exploit spatial context. Although kernel CCA-MAD extends this framework to nonlinear relationships through a reproducing-kernel Hilbert space (RKHS) \cite{nielsen2010linear, Denaro2018}, its practical use for satellite imagery is limited by the need to construct and store kernel matrices whose memory requirement grows quadratically with the number of pixels, making full-scene or large-area processing computationally prohibitive for modern Earth observation data containing millions of samples. Moreover, the nonlinear mapping is predetermined by the selected kernel and remains essentially pixel based. Deep CCA (DCCA) \cite{pmlrandrew13} and Deep GCCA (DGCCA) \cite{benton2019deep} provide a more scalable alternative by replacing linear projections with trainable neural encoders, while DCCAE \cite{pmlrwangb15} further incorporates a reconstruction term to preserve informative representations. For multitemporal PIFs extraction or later called deep-pseudoinvariant features (DIFs), convolutional encoders are particularly attractive because they can learn nonlinear spatial–spectral relationships while exploiting local texture, edges, and neighborhood consistency without requiring explicit pairwise kernel storage.
Although there are more advanced AI models in the remote sensing area, such as Mamba, Transformer, and Quantum Deep Network (QUEEN) \cite{IWDQUEEN,tang2025quantum,HyperQUEEN}, recent remote sensing literature shows that simple convolutional deep network is indeed sufficient to achieve high-performance DIF learning directly in the original optical domain \cite{HyperLUCID}, motivating this paper to develop lightweight deep model for surface reflectance harmonization.

More importantly, although convolutional architectures have been widely adopted for supervised change detection \cite{Daudt2018, Chen2020, Khelifi2020, Shi2020, liu2025cd}, they typically depend on labeled change masks, which are labor-intensive and less practical. In parallel, unsupervised and self-supervised deep change detection has advanced rapidly \cite{Saha2019unsupervised, Chen2022unsupervised, Zhang2023self-supervised, Jian2024uncertainty-aware, Li2025unsupervised}; however, these methods output change maps or change representations rather than an ordered set of invariance-related canonical components, and therefore cannot be directly exploited for pseudoinvariant feature extraction and  harmonization.
Again inspired by the aforementioned work \cite{HyperLUCID}, the DIF can be effectively learned under a fully \textit{unsupervised} manner, by exploiting the so-called incremental safe pseudo-labeling \cite{HyperLUCID}, with successful applications to spectral calibration and change detection even over rather sophisticated land-cover scenarios.
Based on the recent advances \cite{HyperLUCID}, in the present setting of our paper, the proposed learning process aims to be simple, lightweight, and fully \textit{unsupervised}, with multiview correlation itself providing the supervisory signal for invariant feature extraction. The flexibility introduced by nonlinear representation learning also creates a critical risk such as representation collapse. When the multiview correlation objective is optimized without an explicit mechanism that preserves information from the input data, a highly expressive encoder may increase cross-date agreement by mapping different observations to nearly identical or constant representations, thereby suppressing the temporal variability that the method is intended to detect.
This form of degeneracy might be addressed by the so-called safe pseudo-labeling \cite{HyperLUCID}, and is closely related to collapse in joint-embedding self-supervised learning, where deep Siamese architectures \cite{CODEMM} require additional constraints to avoid trivial solutions, such as the stop-gradient and predictor mechanism \cite{Chen2021}, the momentum-updated target network \cite{NEURIPS2020_f3ada80d}, or the decorrelation objective \cite{pmlrzbontar21a}. The latter is particularly relevant to the present formulation because its objective explicitly regulates the cross-correlation structure between learned representations, making it conceptually similar to correlation-driven multiview deep learning.

The principal contribution of this work is not merely to replace the linear projections in GCCA with neural networks, but to develop Deep MGCCA specifically to address the risk of representational collapse inherent in nonlinear multiview correlation learning. The proposed architecture uses shared nonlinear encoder with date-specific nonlinear head outputs, preserving the GCCA structure at the stage where cross-date agreement is enforced. Any structure common to all dates is learned once, in the shared encoder, while the per-date heads account only for residual date-specific behavior. A reconstruction regularizer, inspired by DCCAE \cite{pmlrwangb15}, is incorporated to retain information from the original observations and prevent the encoder from erasing change-sensitive content, while an explicit robust weighting mechanism inherited from IR-MAD distinguishes meaningful multitemporal convergence from degenerate agreement caused by feature collapse. Together, these components yield informative invariant representations rather than trivial cross-date agreement, and the resulting deep-pseudoinvariant features support both principled change detection and harmonization.

Once the DIFs are extracted, a transformation is applied for surface reflectance harmonization. Linear invariant features are typically addressed using linear transformations, whereas nonlinear invariant features require nonlinear transformations \cite{DenaroC2019}. During transformation, preserving spectral similarity is also essential to maintain consistent spectral signatures \cite{SYARIZ201956, denaro2026}. By combining these aspects, the proposed approach applies nonlinear transformation to the extracted DIFs, enabling compact harmonization while preserving spectral signatures that are crucial for downstream remote sensing missions, such as identification and classification \cite{Denaro2020, lin2015identifiability, HyperCSI, PAINT}.

The contributions of this work are summarized as follows:
\begin{enumerate}
\item A convex, globally fitted monotone piecewise-linear harmonization transform, estimated on the quantile functions of the invariant-pixel population of all date pairs, with temporal and spectral consistency preservation, and curvature regularization, generalizing globally optimization without privileging any reference date.
\item DeepMGCCA, an unsupervised deep multiview extension of IR-MAD in which a shared convolutional encoder with date-specific head outputs of features learning, the weighted unit-variance constraint is enforced by construction through weighted center-scaling, and a reconstruction regularizer prevents degenerate collapse of the free nonlinear reparameterizations.
\item A two-direction spectral-angle evaluation protocol spectral angle mapper (SAM) (SAM$_h$ between harmonized dates, SAM$_{sc}$ between each image and its correction) that separates between-date agreement from self-distortion, applied together with RMSE, PSNR, SSIM, and correlation measures.
\item A systematic comparison on four regions of interest spanning urban, desert, coastal, and forest environments with Landsat-8/9 and Sentinel-2 acquisitions, against reference-affine \cite{DU2002123}, sequential stepwise affine \cite{lin2019}, and globally spectral-constrained baselines \cite{denaro2026}.
\end{enumerate}

The remainder of this paper is organized as follows. Section ~\ref{sec:meth} presents the proposed methodology, including (a) the linear GCCA formulation, (b) the nonlinear GCCA formulation, and (c) the nonlinear globally constrained transformation based on the extended formulation of Denaro \cite{denaro2026}. Section ~\ref{sec: experiment} describes the experimental data, implementation settings, evaluation strategy and the discussions. Finally, Section ~\ref{sec:conclusion} summarizes the main findings and outlines directions for future research.

\section{Methodology}\label{sec:meth}

This study develops DeepMGCCA for invariant pixel selection in
multitemporal remote sensing imagery, building on the classical GCCA
framework. The objective is to identify stable and representative pixels
that maintain consistent information across acquisition dates while
minimizing the influence of temporal variations. Section~\ref{sec:linear}
first recalls the linear GCCA--MAD formulation \cite{lin2019, denaro2026}
that serves as the backbone of the proposed framework, together with its
kernel extension (KGCCA), which generalizes classical GCCA to nonlinear
relationships but is impractical at scene scale; the proposed DeepMGCCA
is then developed as the deep learning instantiation of the same
framework, replacing the fixed linear projections with a shared
convolutional encoder while retaining the GCCA correlation structure,
its change statistic, and its iterative reweighting.

\subsection{Linear GCCA approach}\label{sec:linear}

The linear GCCA framework characterizes the relationships among multiple observations under the assumption of linear dependence between their representations \cite{Tenenhaus_Tenenhaus_2011}. Invariant features are identified by solving a joint optimization problem subject to the constraints, formulated as:

\begin{equation}
\begin{aligned}
& \underset{\mathbf{a}_1, \ldots, \mathbf{a}_J}{\text{maximize}} \;\; \sum_{j,k=1;\, j\neq k}^{J} C_{jk} \, \frac{\mathbb{E}\!\left[u_j u_k\right]}{\sqrt{\mathbb{E}\!\left[u_j^2\right]\,\mathbb{E}\!\left[u_k^2\right]}} \\
& \text{subject to} \;\; \mathbb{E}\!\left[u_j^2\right] = \mathbb{E}\!\left[u_k^2\right] = 1,
\end{aligned}
\label{eq:gcca}
\end{equation}
where $u_j = \mathbf{a}_j^T\mathbf{X}_j$ and $u_k = \mathbf{a}_k^T\mathbf{X}_k$, and $\mathbf{X}_j$, $\mathbf{X}_k$ are assumed mean-centered so that $\mathbb{E}[u_j u_k] = \mathbf{a}_j^T\,\mathbb{E}[\mathbf{X}_j\mathbf{X}_k^T]\,\mathbf{a}_k$ coincides with the covariance $\mathbf{a}_j^T\Sigma_{jk}\mathbf{a}_k$; and $C_{jk}\in\{0,1\}$ is the design matrix of the criterion, with $C_{jk}=1$ when dates $j$ and $k$ are connected and $C_{jk}=0$ otherwise.

Since $\mathbf{a}_j^T\Sigma_{jj}\mathbf{a}_j = \mathbf{a}_k^T\Sigma_{kk}\mathbf{a}_k = 1$, the variance of the difference variate $u_j - u_k$ reduces to
\begin{equation}
\begin{aligned}
\mathrm{Var}(u_j - u_k) &= \mathbb{E}\!\left[u_j^2\right] + \mathbb{E}\!\left[u_k^2\right] - 2\,\mathbb{E}\!\left[u_j u_k\right] \\
&= 2 - 2\,\mathbb{E}\!\left[u_j u_k\right].
\end{aligned}
\label{eq:mad-var}
\end{equation}
Equation~\eqref{eq:mad-var} shows that maximizing $\mathbb{E}[u_j u_k]$ in \eqref{eq:gcca} is equivalent to \emph{minimizing} $\mathrm{Var}(u_j-u_k)$, i.e., the variance of the difference variate $u_j - u_k = \mathbf{a}_j^T\mathbf{X}_j - \mathbf{a}_k^T\mathbf{X}_k$, which is precisely the MAD variate used to construct $\mathbf{MAD}_j$ in \eqref{eq:mad} \cite{CANTY20081025, CANTY2004441, NIELSEN19981}.This relationship defines the linear combination over the set of connected images, as expressed in
\begin{equation}
\mathbf{MAD}_j = \sum_{j,k=1;\, k\neq j}^{J} {C_{jk} \boldsymbol{\rho}_{jk}   \mathbf{MAD}_{jk}},
\label{eq:mad}
\end{equation}
where $\rho_{jk}$ denotes the correlation between the transformed variates $u_j$ and $u_k$ of the connected image pair $(\mathbf{X}_j, \mathbf{X}_k)$, obtained from the solution of \eqref{eq:gcca}, i.e., $\rho_{jk} = \mathbb{E}[u_j u_k]$, and $\mathbf{MAD}_j = [\mathbf{mad}_1 \; \ldots \; \mathbf{mad}_B]^T \in \mathbb{R}^{B\times n}$, where $n$ number of pixels . The NMAD image corresponding to $\mathbf{X}_j$ is then obtained from the sum of squares of the standardized MAD variates:
\begin{equation}
\mathbf{nmad}_j = \sum_{b=1}^{B} \left( \frac{\mathbf{mad}_{b,j}}{\sigma_{\mathbf{mad}_b}} \right)^{2} \in \chi^2(B), \quad j = 1, \ldots, J,
\label{eq:nmad}
\end{equation}
where $\mathbf{mad}_{b,j}$ denotes the $b$-th band component of $\mathbf{MAD}_j$ (i.e., the $b$-th entry of $[\mathbf{mad}_1 \; \ldots \; \mathbf{mad}_B]^T$ associated with image $j$), $\sigma_{\mathbf{mad}_b}$ is the standard deviation of the $b$-th band's MAD variate, and $B$ is the number of spectral bands, which also serves as the degrees of freedom of the resulting chi-square distribution. Once the combined NMAD images are obtained, a threshold is applied to convert the resulting probabilities into a binary change map $\{0,1\}$, separating pseudo-invariant features (PIFs) from non-PIFs within $\mathbf{X}_j$.

To down-weight changed pixels, the population moments $\mathbb{E}[u_ju_k]$, $\mathbb{E}[u_j^2]$, $\mathbb{E}[u_k^2]$ in \eqref{eq:gcca} are replaced by weighted sample estimates, using pixel weights $w_\ell^{(t)}\in[0,1]$, $\ell=1,\ldots,n$, updated over iterations $t$:
\begin{equation}
\begin{aligned}
\bar{\mathbf{X}}_j^{(t)} &= \frac{\sum_{\ell} w_\ell^{(t)} \mathbf{X}_{j,\ell}}{\sum_{\ell} w_\ell^{(t)}}, \\[4pt]
\Sigma_{jk}^{(t)} &= \frac{\sum_{\ell} w_\ell^{(t)} \left(\mathbf{X}_{j,\ell}-\bar{\mathbf{X}}_j^{(t)}\right)\left(\mathbf{X}_{k,\ell}-\bar{\mathbf{X}}_k^{(t)}\right)^T}{\sum_{\ell} w_\ell^{(t)}},
\end{aligned}
\label{eq:wstats}
\end{equation}
so that $\mathbb{E}[u_ju_k] \approx \mathbf{a}_j^T\Sigma_{jk}^{(t)}\mathbf{a}_k$ and $\mathbb{E}[u_j^2] \approx \mathbf{a}_j^T\Sigma_{jj}^{(t)}\mathbf{a}_j$ at iteration $t$. At $t=0$, weights are seeded from the spectral angle $\theta_{jk,\ell}=\arccos\!\left(\frac{\mathbf{X}_{j,\ell}^T\mathbf{X}_{k,\ell}}{\lVert\mathbf{X}_{j,\ell}\rVert\lVert\mathbf{X}_{k,\ell}\rVert}\right)$, $w_\ell^{(0)}=1-\theta_{jk,\ell}/(\pi/2)$, with $\bar{\mathbf{X}}_j^{(0)}$ the unweighted temporal mean. For $t\geq1$,
\begin{equation}
w_\ell^{(t)} = 1 - F_{\chi^2(B)}\!\left(\mathrm{nmad}_{j,\ell}^{(t-1)}\right),
\label{eq:inv-nmad}
\end{equation}
and the process repeats through \eqref{eq:gcca}--\eqref{eq:nmad} until convergence.

\subsection{Nonlinear GCCA approach}\label{sec:nonlinear}

Unlike linear GCCA, nonlinear GCCA aims to capture higher-order dependencies among multitemporal observations by mapping the original data into a nonlinear feature space. Two nonlinear strategies are investigated in this research: kernel-based GCCA and deep learning-based GCCA.

\begin{figure}[t]
\centering
\includegraphics[width=1\linewidth]{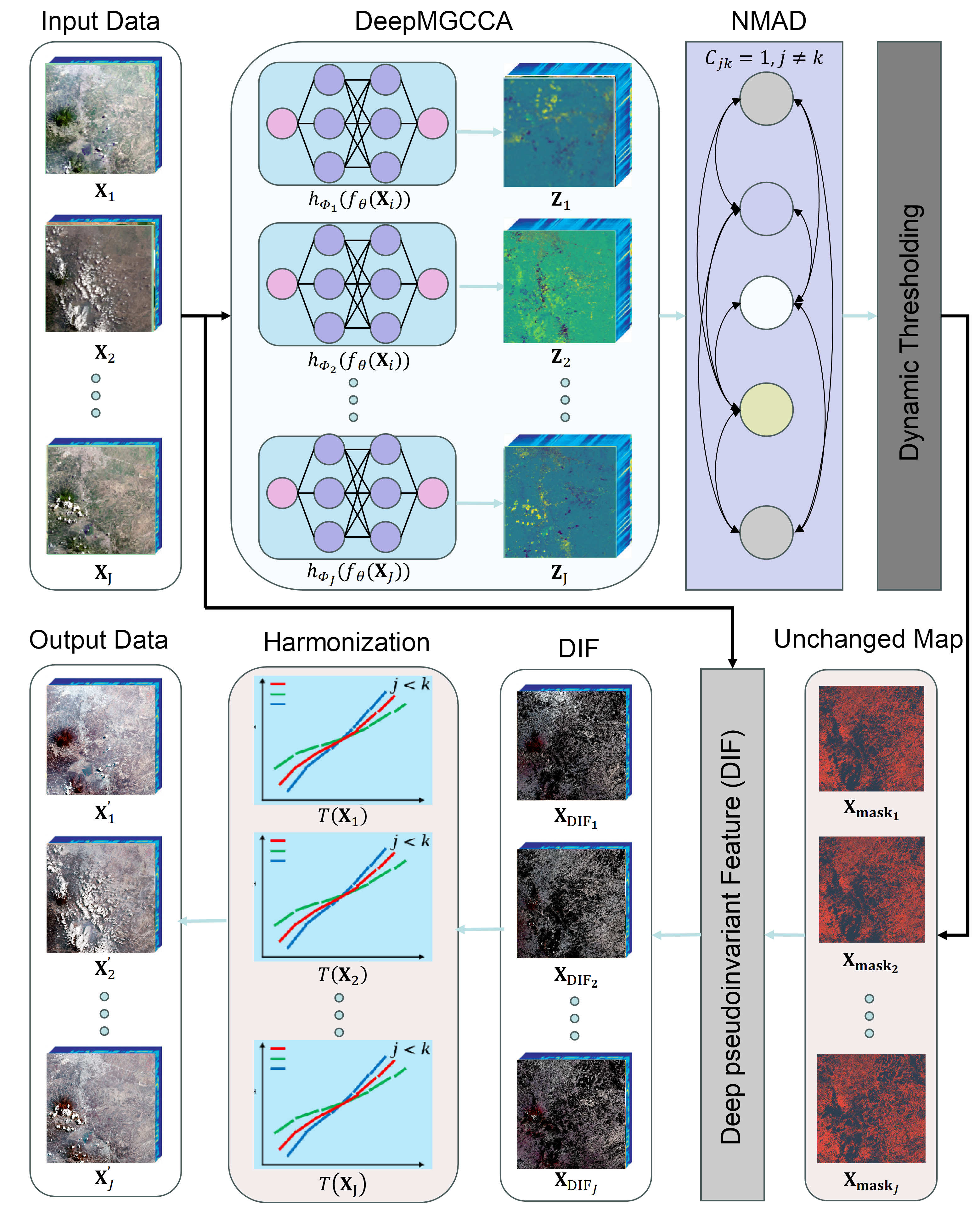}
\caption{Overview of the proposed DeepMGCCA framework and harmonization pipeline. \textit{Top row:} each input image $\mathbf{X}_j$, $j=1,\ldots,J$, is mapped to a learned feature representation $\mathbf{Z}_j = h_{\phi_j}(f_\theta(\mathbf{X}_j))$ through the shared encoder $f_\theta$ followed by a date-specific head output $h_{\phi_j}$, trained end-to-end by backpropagating the multiview correlation and reconstruction objectives. The consensus-aligned canonical variates $\mathbf{U}_j$ are then differenced across the connected date pairs ($C_{jk}=1$, $j\neq k$) and aggregated into the normalized MAD statistic $\mathrm{NMAD}_j$. \textit{Bottom row:} a dynamic threshold converts $\mathrm{NMAD}_j$ into the unchanged map $\mathbf{X}_{\text{mask}_{j}}$, from which the DIF of all dates are extracted $\mathbf{X}_{\text{DIF}_{j}}$. The DIF quantiles determine the knot grid of a monotone piecewise-linear transform $\mathrm{T}(\cdot)$, one per date and fitted over all pairs $j<k$, which is applied to every pixel of the scene to produce the harmonized images $\mathbf{X}'_1,\ldots,\mathbf{X}'_J$.}
\label{fig:general}
\end{figure}

\subsubsection{Kernel approach} The kernel GCCA approach extends the linear formulation in \eqref{eq:gcca} by replacing each linear projection $u_j = \mathbf{a}_j^T\mathbf{X}_j$ with a projection in a reproducing kernel Hilbert space, $u_j = \mathbf{a}_j^T\phi_j(\mathbf{X}_j)$, where $\phi_j(\cdot)$ is an implicit nonlinear feature map associated with a kernel function $k_j(\mathbf{X}_{j,\ell}, \mathbf{X}_{j,\ell'}) = \phi_j(\mathbf{X}_{j,\ell})^T\phi_j(\mathbf{X}_{j,\ell'})$\cite{Denaro2018, TENENHAUS2015114}. By the representer theorem, $\mathbf{a}_j$ is expressed as a linear combination of the mapped pixels, $\mathbf{a}_j = \sum_{\ell=1}^n \alpha_{j,\ell}\,\phi_j(\mathbf{X}_{j,\ell})$, so that \eqref{eq:gcca} is rewritten entirely in terms of the $n\times n$ kernel (Gram) matrices $\mathbf{K}_j$, with entries $[\mathbf{K}_j]_{\ell,\ell'} = k_j(\mathbf{X}_{j,\ell}, \mathbf{X}_{j,\ell'})$, in place of $\Sigma_{jk}$:
\begin{equation}
\begin{aligned}
& \underset{\boldsymbol{\alpha}_1, \ldots, \boldsymbol{\alpha}_J}{\text{maximize}} \;\; \sum_{j,k=1;\, j\neq k}^{J} C_{jk} \, \frac{\boldsymbol{\alpha}_j^T \mathbf{K}_j \mathbf{K}_k \boldsymbol{\alpha}_k}{\sqrt{\boldsymbol{\alpha}_j^T \mathbf{K}_j^2 \boldsymbol{\alpha}_j \; \boldsymbol{\alpha}_k^T \mathbf{K}_k^2 \boldsymbol{\alpha}_k}} \\
& \text{subject to} \;\; \boldsymbol{\alpha}_j^T \mathbf{K}_j^2 \boldsymbol{\alpha}_j = \boldsymbol{\alpha}_k^T \mathbf{K}_k^2 \boldsymbol{\alpha}_k = 1,
\end{aligned}
\label{eq:kgcca}
\end{equation}
where a ridge term $\eta\mathbf{K}_j$ ($\eta > 0$) is typically added to $\mathbf{K}_j^2$ to regularize the otherwise ill-conditioned inversion. This substitution allows GCCA to model nonlinear correlations while preserving the generalized eigenvalue structure of the original optimization problem, without requiring $\phi_j(\cdot)$ to be evaluated explicitly.

In principle, \eqref{eq:kgcca} enables kernel GCCA to extract nonlinear invariant features that linear projections in \eqref{eq:gcca} cannot capture, using kernel functions suited to complex spectral relationships. In practice, however, this formulation is largely impractical for multitemporal remote sensing imagery. The kernel matrix $\mathbf{K}_j$ scales as $n \times n$ in the number of pixels $n$, so a single moderately sized scene ($n$ on the order of $10^6$ pixels or more) already yields a kernel matrix with $10^{12}$ entries, exceeding the memory of ordinary computing hardware, while solving \eqref{eq:kgcca} requires operations on $\mathbf{K}_j$ that scale as $O(n^3)$ in time and $O(n^2)$ in memory. This cost is further compounded under the iterative reweighting scheme in \eqref{eq:wstats}: since $\mathbf{K}_j$ depends on the full pixel set, the entire kernel matrix must be recomputed at every iteration $t$, multiplying an already prohibitive cost by the number of iterations to convergence. For these reasons, kernel-based GCCA is reported here primarily as a baseline for comparison rather than a scalable solution, motivating the deep learning-based approach described next.

\begin{figure*}[t]
\centering
\includegraphics[width=0.9\linewidth]{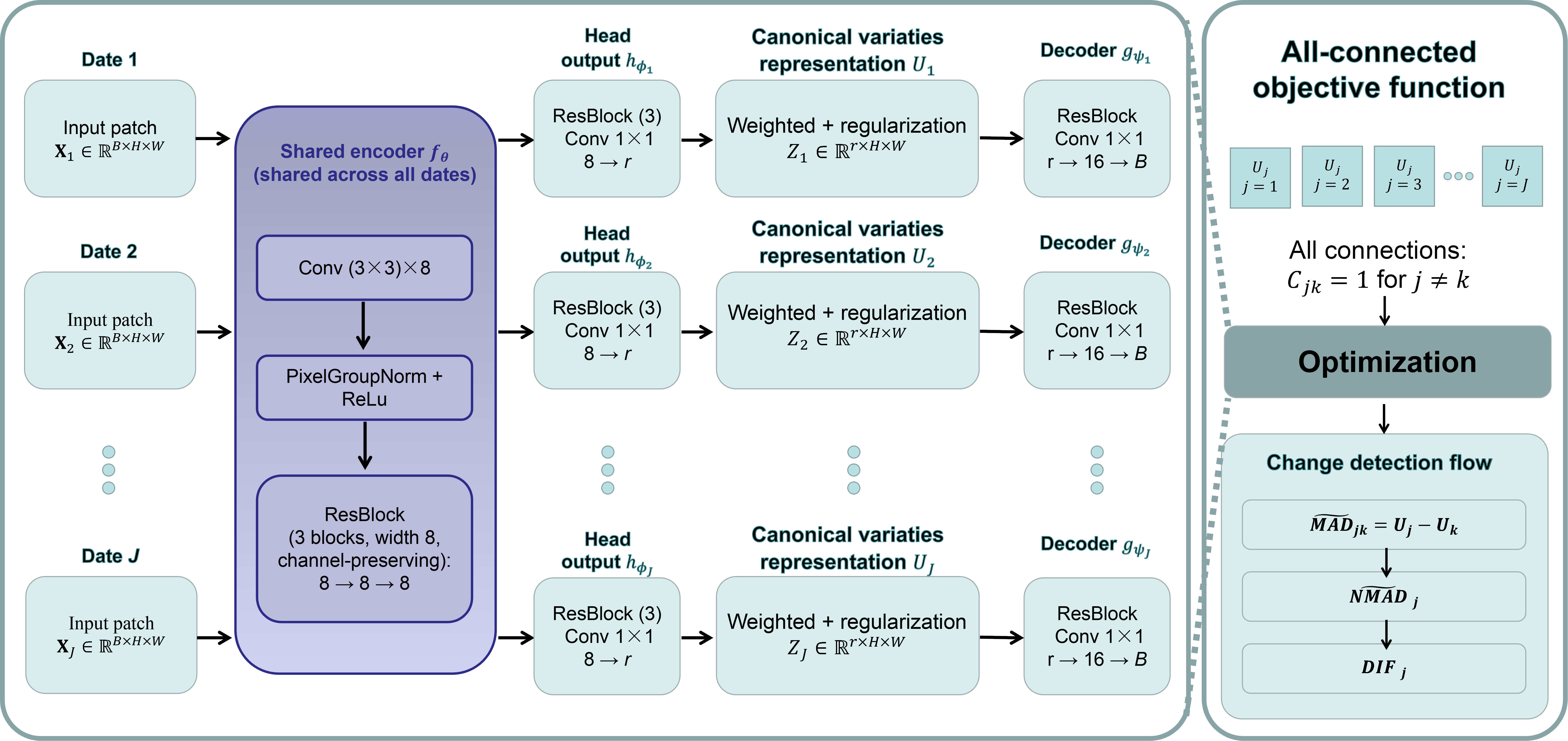}
\caption{DeepMGCCA architecture. \textit{Left:} for each date
$j=1,\ldots,J$, the input patch $\mathbf{X}_j\in\mathbb{R}^{B\times H\times W}$ passes through a shared nonlinear encoder $f_\theta$ with weights tied across all $J$ dates, followed by a date-specific head $h_{\phi_j}$ that produces the head outputs $\mathbf{Z}_j = h_{\phi_j}(f_\theta(\mathbf{X}_j)) \in \mathbb{R}^{r\times H\times W}$, from which the canonical variates $\mathbf{U}_j$ are formed by consensus alignment after training. The weighted unit-variance constraint is enforced by weighted center-scaling, and a date-specific decoder $g_{\psi_j}$ reconstructs $\mathbf{X}_j$ from $\mathbf{Z}_j$, contributing the reconstruction regularizer. \textit{Right:} the multiview correlation objective and the reconstruction regularizer are jointly optimized over $\theta$, $\{\phi_j\}_{j=1}^J$, and $\{\psi_j\}_{j=1}^J$. The trained canonical variates $\mathbf{U}_j$ are fed into the change-detection flow: pairwise difference variates $\widetilde{\mathbf{MAD}}_{jk} = \mathbf{U}_j - \mathbf{U}_k$ are aggregated into the normalized MAD (NMAD) statistic and thresholded to produce the final deep-pesudoinvariant-feature (DIF) image.}
\label{fig:arch}
\end{figure*}

\subsubsection{Deep learning approach} Deep-learning-based GCCA replaces each linear projection
$u_j = \mathbf{a}_j^T\mathbf{X}_j$ in \eqref{eq:gcca} with a nonlinear one, learned end-to-end rather than solved in closed form, as illustrated in Fig.~\ref{fig:general}. A single shared encoder $f_\theta$ is applied identically to every date and composed with a date-specific head $h_{\phi_j}$ (Fig.~\ref{fig:general}, left) to produce the head outputs, i.e., the learned feature representations,
\begin{equation}
\mathbf{Z}_j = h_{\phi_j}\!\left(f_\theta(\mathbf{X}_j)\right) \in \mathbb{R}^{r\times H\times W}, \quad j = 1,\ldots,J,
\label{eq:canonical-variates}
\end{equation}
which generalize $u_j$ of \eqref{eq:gcca} from a scalar per-pixel projection to an $r$-dimensional one. The canonical variates proper, $\mathbf{U}_j$, are then formed from $\mathbf{Z}_j$ by consensus alignment, \eqref{eq:aligned-variates}, once training is complete. Any spectral-spatial structure shared across dates is thus learned once, in $\theta$, rather than separately per date; only $\{\phi_j\}_{j=1}^J$ differ across dates, accounting for residual date-specific behavior. In this work $f_\theta$ is a convolutional encoder, so that $\mathbf{Z}_j$ exploits local spatial context in addition to spectral information. Writing $\bar{\mathbf{Z}}_j\in\mathbb{R}^{n\times r}$ ($n=H\times W$) for the flattened, pixel-weighted, mean-centered head outputs under the weight matrix $D_w=\mathrm{diag}(w_1,\ldots,w_n)$ from \eqref{eq:inv-nmad}, the encoder and heads are trained so that each of them satisfies the weighted unit-variance constraint,
\begin{equation}
\bar{\mathbf{Z}}_j^T D_w \bar{\mathbf{Z}}_j = \mathbf{I}_r, \quad j = 1,\ldots,J,
\label{eq:weighted-constraint}
\end{equation}
the nonlinear counterpart of $\mathbf{a}_j^T\Sigma_{jj}\mathbf{a}_j=1$ in \eqref{eq:gcca}.
Since every $\bar{\mathbf{Z}}_j$ is weighted-center-scaled at each training step before the criterion is evaluated. The network parameters are learned by

\begin{equation}
\begin{split}
\theta^*, \{\phi_j^*\}, \{\psi_j^*\} =
\underset{\theta, \phi, \psi}{\arg\max} \;\;
& \mathcal{L}_{\text{dmgcca}}\big(\{\bar{\mathbf{Z}}_j(\theta,\phi_j)\}\big) \\
& - \lambda_{\mathrm{recon}}\,\mathcal{R}\big(\{\mathbf{Z}_j(\theta,\phi_j)\},\{\psi_j\}\big),
\end{split}
\label{eq:deep-gcca-objective}
\end{equation}

\begin{equation}
\mathcal{R} = \sum_{j=1}^{J} \left\lVert
\mathbf{X}_j - g_{\psi_j}\!\big(\mathbf{Z}_j\big) \right\rVert_F^2,
\label{eq:recon}
\end{equation}
%
where $\mathcal{L}_{\text{dmgcca}}$, DeepMGCCA objective function, uses the criteria for the maximum-variance optimization\cite{benton2019deep}, \eqref{eq:maxvar}, evaluated on the weighted, centered
head outputs $\bar{\mathbf{Z}}_{j}$. $\mathcal{R}$ in
\eqref{eq:recon} is the total reconstruction error, the sum over dates of
the squared Frobenius distance between each input image $\mathbf{X}_j$
and the image rebuilt from its head output by the decoder
$g_{\psi_j}$, which the $\mathbf{Z}_{j}$ retain
enough information to reconstruct the inputs, and is measured against a
constant-predictor baseline of order unity. $\lambda_{\mathrm{recon}}>0$ weights this regularizer against the
correlation criterion. Since the date-specific heads are free nonlinear
reparameterizations, the correlation criterion alone admits trivial
maximizers that erase information. Eq. \eqref{eq:recon} forces each
$\mathbf{Z}_j$ to remain decodable to $\mathbf{X}_j$, preventing
degenerate collapse while leaving the correlation objective itself
untouched.

Under the criterion \cite{Horst_1961,benton2019deep}, a shared consensus $\mathbf{G}\in\mathbb{R}^{n\times r}$ and per-date linear maps $\mathbf{A}_j\in\mathbb{R}^{r\times r}$ are found jointly with $\theta,\{\phi_j\}$ by
\begin{equation}
\begin{aligned}
\mathcal{L}_{\text{dmgcca}}
&= -\min_{\mathbf{G},\,\mathbf{A}_1,\ldots,\mathbf{A}_J}
\sum_{j=1}^{J}
\left\lVert \mathbf{G} - \bar{\mathbf{Z}}_j\mathbf{A}_j \right\rVert_F^2,
\\
&\quad \text{s.t.} \quad
\mathbf{G}^T\mathbf{G} = \mathbf{I}_r.
\end{aligned}
\label{eq:maxvar}
\end{equation}
solved in closed form at each training step via the $(Jr\times Jr)$ eigenproblem $\mathbf{K}=\mathbf{H}^T\mathbf{H}$, $\mathbf{H}_j=\bar{\mathbf{Z}}_j\big(\bar{\mathbf{Z}}_j^T\bar{\mathbf{Z}}_j\big)^{-1/2}$, so only $\theta,\{\phi_j\}$ carry learned gradients. Two properties motivate this choice: the consensus $\mathbf{G}$ fixes the cross-date component correspondence during training, so the difference variates below are meaningful by construction; and the constraint \eqref{eq:weighted-constraint} is enforced exactly by the weighted center-scaling. Once trained, the consensus-aligned variates are
\begin{equation}
\mathbf{U}_j = \bar{\mathbf{Z}}_j \mathbf{A}_j, \quad j = 1,\ldots,J,
\label{eq:aligned-variates}
\end{equation}
and the invariant features flow (Fig.~\ref{fig:general}, right) follows \eqref{eq:mad}--\eqref{eq:nmad} with $\mathbf{U}_j$ in place of $u_j$: pairwise difference variates
\begin{equation}
\widetilde{\mathbf{MAD}}_{jk} = \mathbf{U}_j - \mathbf{U}_k, \quad (j,k) \in \mathcal{P},
\label{eq:mad-deep}
\end{equation}
are aggregated per date, standardized into $\mathrm{nmad}_j$ as in \eqref{eq:nmad}, and thresholded into the final DIFs image. The complete training procedure is given in Algorithm~\ref{alg:deepgcca}.

The deep framework uses the same iterative reweighting scheme \eqref{eq:inv-nmad} as the linear framework, applied here to the projected spaces $\mathbf{Z}_j$. Its first iteration requires an initial pixel weight $w^{(0)}$. Rather than being computed on the raw reflectance, $w^{(0)}$ is built directly from the projected multitemporal scenes $\mathbf{Z}_1,\ldots,\mathbf{Z}_J$, obtained from an initial forward pass of the encoder and heads, by combining a \emph{temporal} factor with a \emph{latent spectral-angle} factor. First, a cloud-robust temporal reference is formed per latent channel and per pixel by taking the lower of the mean and the median across the $J$ acquisition dates:
\begin{equation}
m_{{T}_{r,\ell}} = \min\!\Big( \underbrace{\tfrac{1}{J}\textstyle\sum_{j=1}^{J} Z_{j,r,\ell}}_{\text{temporal mean}},\; \underbrace{\operatorname{median}(Z_{1,r,\ell}, \cdots, Z_{J,r,\ell}) }_{\text{temporal median}} \Big),
\label{eq:tfactor}
\end{equation}
where the statistics are taken over dates only, so $m\_T \in \mathbb{R}^{r\times H\times W}$ retains the full latent-channel and spatial dimensions. A bright outlier on most of the dates (e.g., cloud) inflates the mean but not the median, and the $\min(\cdot)$ keeps the robust estimate; a pixel whose $\mathbf{Z}_{j,\ell}$ stays close to this reference on every date is invariant. Second, the Spectral Angle Mapper (SAM) \cite{CODEHCD, lin2019, denaro2026} is computed per-date rather than per-pair on the feature spaces: for each ordered pair $C_{j,k}$ connection, $j\neq k$,
\begin{equation}
\begin{aligned}
\theta_{jk,\ell} &= \arccos\!\left(\frac{\mathbf{Z}_{j,\ell}^T\mathbf{Z}_{k,\ell}}{\lVert\mathbf{Z}_{j,\ell}\rVert\,\lVert\mathbf{Z}_{k,\ell}\rVert}\right), \\
\rho_{jk,\ell} &= \max\!\big(0,\; \cos\theta_{jk,\ell}\big),
\end{aligned}
\label{eq:sam-pair}
\end{equation}
and the per-date latent SAM score aggregates all partners, scaled by the pairwise correlation so that small, well-correlated angles are trusted more:
\begin{equation}
\mathrm{SAM}_{j,\ell} = \frac{1}{J-1}\sum_{k\neq j} \rho_{jk,\ell}\,\theta_{jk,\ell}.
\label{eq:sam-init}
\end{equation}
The initialization weight is the product of the temporal and spectral-angle factors, each normalized to $[0,1]$ over the valid pixels:
\begin{equation}
w_{j,k,\ell}^{(0)} \;\propto\; \mathrm{SAM}_{j,\ell}\cdot T_{k,\ell},
\qquad w_{j,k,\ell}^{(0)}\in[0,1],
\label{eq:w0}
\end{equation}
and the temporal centering at $t{=}0$ uses the corresponding time-weighted mean, per latent channel, with the temporal factor $m_{T_{k,\ell}}$ already absorbed into $w^{(0)}_{j,k,\ell}$ via \eqref{eq:w0},
\begin{equation}
\bar{Z}_{j,k}^{(0)} = \frac{\sum_{\ell} w_{j,k,\ell}^{(0)}\,Z_{j,k,\ell}}{\sum_{\ell} w_{j,k,\ell}^{(0)}},
\label{eq:fwmean}
\end{equation}
so that pixels already discordant at initialization are down-weighted before the first encoder-training epoch. Thereafter, weights are updated by the same NMAD rule \eqref{eq:inv-nmad} as in the linear framework, applied to the trained canonical variates $\mathbf{U}_j$.

\begin{algorithm}[t]
\caption{DeepMGCCA Feature Learning}
\begin{algorithmic}[1]\label{alg:deepgcca}
\STATE {\bf Given} patches $\mathbf{X}_1,\ldots,\mathbf{X}_J$ ($B$ bands); component count $r$; $f_\theta$; $\{h_{\phi_j}\}$; $\{g_{\psi_j}\}$; $\lambda_{\mathrm{recon}}>0$.
\STATE Initialize $t:=0$; $\theta,\{\phi_j\},\{\psi_j\}$; \; \\
$\mathbf{Z}_j^{(0)} \leftarrow h_{\phi_j}(f_\theta(\mathbf{X}_j))$, $\forall j$ \COMMENT{initial forward pass for weight initialization}
\STATE $w^{(0)}$ from Eqs.~\eqref{eq:tfactor}--\eqref{eq:w0}, computed on $\{\mathbf{Z}_j^{(0)}\}$.
\REPEAT
\STATE $D_w := \mathrm{diag}(w^{(t)})$.
\REPEAT
\STATE $\mathbf{Z}_j = h_{\phi_j}(f_\theta(\mathbf{X}_j))$; \; \\
$\bar{\mathbf{Z}}_j = \text{weighted-center-scale}(\mathbf{Z}_j, D_w)$, $\forall j$
\STATE $\mathbf{H}_j \leftarrow \bar{\mathbf{Z}}_j(\bar{\mathbf{Z}}_j^T\bar{\mathbf{Z}}_j)^{-1/2}$; \; \\
$\mathbf{G} \leftarrow$ top-$r$ eigenvectors of $\mathbf{H}\mathbf{H}^T$; \; \\
$\mathbf{A}_j \leftarrow \arg\min_{\mathbf{A}}\lVert\mathbf{G}-\bar{\mathbf{Z}}_j\mathbf{A}\rVert_F^2$, $\forall j$
\STATE $\mathcal{L}_{\mathrm{total}} = -\sum_{j}\lVert \mathbf{G}-\bar{\mathbf{Z}}_j\mathbf{A}_j\rVert_F^2 - \lambda_{\mathrm{recon}}\sum_{j}\lVert \mathbf{X}_j - g_{\psi_j}(\mathbf{Z}_j)\rVert_F^2$ \COMMENT{Eqs.~\eqref{eq:maxvar}, \eqref{eq:recon}}
\STATE $\theta,\{\phi_j\},\{\psi_j\} \leftarrow \cdot + \eta\,\nabla\mathcal{L}_{\mathrm{total}}$
\UNTIL encoder training converges.
\STATE $\mathbf{U}_j \leftarrow \bar{\mathbf{Z}}_j\mathbf{A}_j$, $\forall j$ \COMMENT{Eq.~\eqref{eq:aligned-variates}}
\STATE $\widetilde{\mathbf{MAD}}_{jk}^{(t)} = \mathbf{U}_j - \mathbf{U}_k$ \COMMENT{Eq.~\eqref{eq:mad-deep}}
\STATE $w^{(t+1)} = 1 - F_{\chi^2(r)}(\mathrm{nmad}_j^{(t)})$; \; \\
$t:=t+1$ \COMMENT{Eqs.~\eqref{eq:nmad}, \eqref{eq:inv-nmad}}
\UNTIL pixel reweighting converges.
\STATE {\bf Output} $\{\mathbf{U}_j\}_{j=1}^J$; $\mathrm{DIF}$.
\end{algorithmic}
\end{algorithm}

\subsection{Image harmonization}\label{sec:spatial}

\begin{figure}[t]
\centering
\includegraphics[width=1\linewidth]{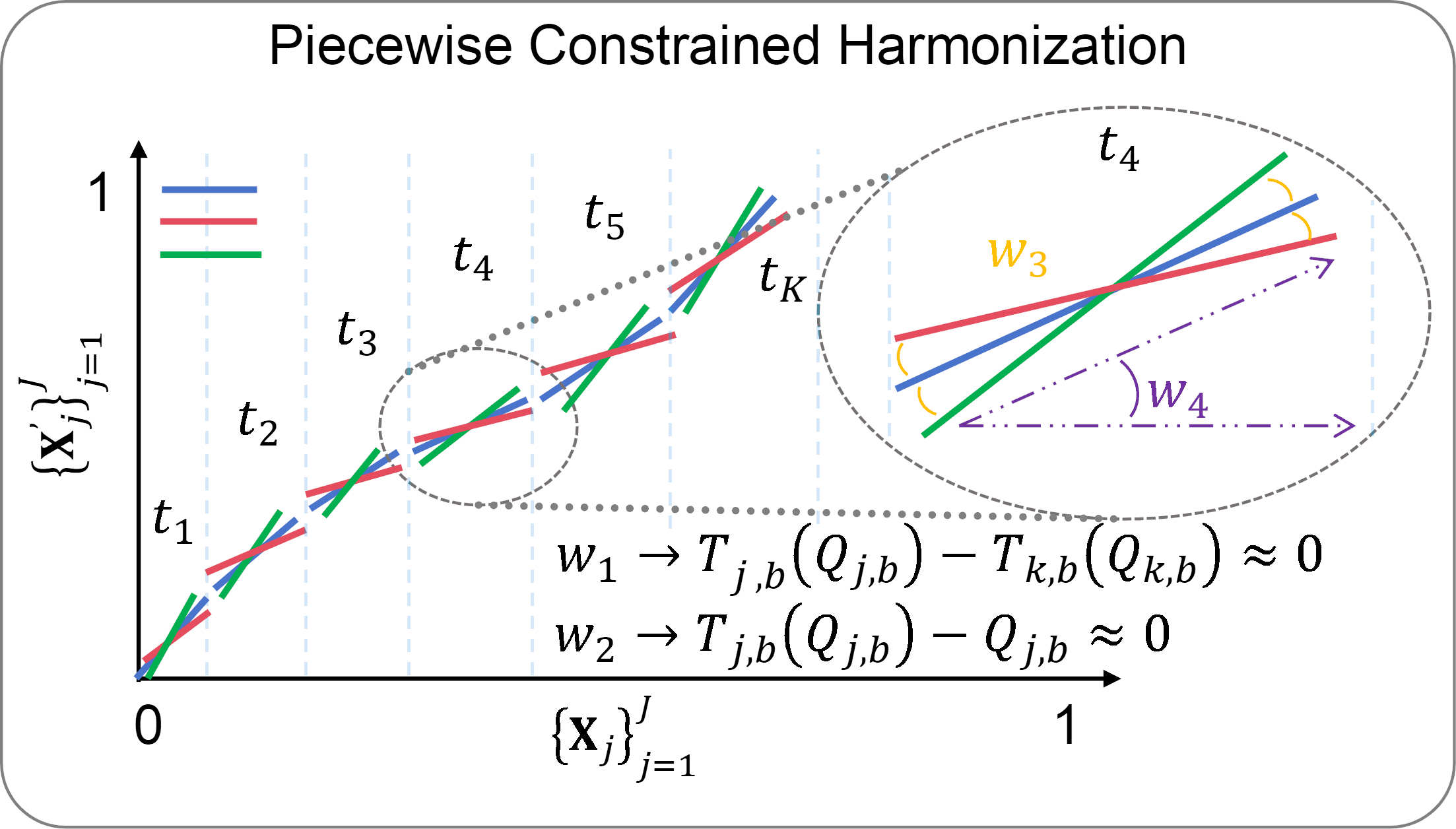}
\caption{Piecewise constrained harmonization. Each date $j$ (colored curves) is mapped by a monotone piecewise-linear transform $T_{j,b}$ defined on a knot grid $t_{b,0}<\cdots<t_{b,K}$ (dashed vertical lines) placed at the quantiles of the pooled DIF population. The agreement term $w_1$ ($E_C$, Eq.~\eqref{eq:11}) drives the transformed quantile functions of all date pairs toward a common level. The anchor term $w_2$ ($E_T$) pulls each transform toward the identity map (dotted diagonal). The band-smoothness term $w_3$ ($E_S$) couples the hinge coefficients across adjacent bands, and and the curvature term $w_4$ ($E_K$, dashed ellipses) penalizes kinks between consecutive knots. Monotonicity is imposed as a box constraint on the hinge coefficients (Eq.~\eqref{eq:10}).}
\label{fig:harmon}
\end{figure}

Once high-quality  DIFs have been extracted, the transformation between each multitemporal image and a common level is determined through constrained harmonization, formulated as a convex, bound-constrained least-squares problem rather than a global affine fit. Fig.~\ref{fig:harmon} illustrates the incorporation of the agreement, anchor, and spectral constraints into this system.

Let $X_{j,b,\ell}$ denote the reflectance of DIF pixel $\ell$ in band $b$ of
date $j$, for $j=1,\ldots,J$ dates and $b=1,\ldots,B$ bands.
Each date-band pair is assigned its own monotone \emph{piecewise-linear} transform $T_{j,b}$, parameterized on a per-band knot grid $t_{b,0}<t_{b,1}<\cdots<t_{b,K}$ (placed at the quantiles of the pooled DIF population, so the knots sit where the data are) via a hinge (ramp) basis:
\begin{equation}
\begin{aligned}
T_{j,b}(x) &= c_{j,b,0} + \sum_{m=1}^{K} c_{j,b,m}\,R_m(x;\, \mathbf{t}_b), \\
R_m(x) &= \mathrm{clip}\!\left(\frac{x - t_{b,m-1}}{t_{b,m}-t_{b,m-1}},\, 0,\, 1\right),
\end{aligned}
\label{eq:9}
\end{equation}
where each $R_m$ is a non-decreasing ramp over the $m$-th knot interval.
This replaces the two-parameter affine transform
$x_j' = x_j\times\alpha + \beta$ used previously with a
$(K{+}1)$-parameter transform per date and band ($K=3$ knots in our
implementation), capable of representing reflectance relationships that
are consistent but not strictly linear across the sensor's dynamic
range, while remaining exactly piecewise linear, hence stable to
apply and trivial to audit. Since every ramp is non-decreasing,
monotonicity of $T_{j,b}$ reduces to a box constraint on its
coefficients rather than a general inequality:
\begin{equation}
\begin{aligned}
c_{j,b,m} &\geq \epsilon_b\, h_{b,m}, \quad m=1,\ldots,K, \\
h_{b,m} &= t_{b,m}-t_{b,m-1}, \quad \epsilon_b = \text{min}_\text{slope},
\end{aligned}
\label{eq:10}
\end{equation}
i.e., the slope of every segment, $T'_{j,b}=c_{j,b,m}/h_{b,m}$ on
interval $m$, is bounded below by $\text{min}_\text{slope}$. A strictly positive bound
forbids flat segments, which would compress bright pixels onto a single
value. The first and last ramps are left open-ended so that pixels
outside the DIF range are linearly extrapolated with the end slopes
rather than collapsed onto the end knots.

Because the DIF pixels are not spatially paired across dates in a way
that supports a per-pixel consensus, the agreement term is stated on the
\emph{quantile functions} of the DIF population. Let $Q_{j,b}(q)$ denote
the $q$-quantile of date $j$, band $b$, over the common DIF set, at $M$
levels ($M=41$, $q\in[0.01,0.99]$). Two dates agree on the invariant set
iff their transformed quantile functions coincide, so the agreement and
anchor terms are
\begin{equation}
\begin{aligned}
E_{C} &= \!\!\sum_{j<k}\sum_{b}\sum_{q} \left[T_{j,b}\!\left(Q_{j,b}(q)\right) - T_{k,b}\!\left(Q_{k,b}(q)\right)\right]^2, \\
E_{T} &= \sum_{j,b,q}\left[T_{j,b}\!\left(Q_{j,b}(q)\right) - Q_{j,b}(q)\right]^2 .
\end{aligned}
\label{eq:11}
\end{equation}
$E_C$ penalizes disagreement between every pair of dates at every
quantile level of the affine formulation \cite{denaro2026}
is exactly its two-parameter special case, recovered at $K=1$. $E_T$
anchors every date's transform toward the identity map on its own
quantile function. No date is privileged as a reference, and the common
level is found globally by the agreement term, while the anchor
prevents the ensemble from sliding jointly to a common but arbitrary
level.

The spectral and curvature terms are
\begin{equation}
\begin{aligned}
E_{S} &= \sum_{j,b,m} \left[c_{j,b+1,m} - c_{j,b,m}\right]^2, \\
E_{K} &= \sum_{j,b,m} \left[c_{j,b,m+1} - c_{j,b,m}\right]^2 .
\end{aligned}
\label{eq:12}
\end{equation}
$E_S$ constrains corresponding hinge coefficients to change smoothly
from band $b$ to band $b{+}1$, generalizing the earlier
second-difference constraint on $\alpha(\lambda_j),\beta(\lambda_j)$ to
the $(K{+}1)$-coefficient case, and so preserves the spectral signature
across bands. $E_K$ penalizes the second difference of the coefficients
\emph{across knots}, the discrete curvature of $T_{j,b}$ itself,
keeping each transform smooth rather than kinked where adjacent
quantile rows pull in different directions. This term has no counterpart
in the affine formulation, whose two-parameter line has no curvature to
constrain.

The four terms combine into the constrained harmonization objective
\begin{equation}
\begin{aligned}
F(c) = \min_{c}\;\; & w_1 E_C + w_2 E_T + w_3 E_S + w_4 E_K, \\
\text{s.t.} \;\; & c_{j,b,m} \geq \epsilon_b\, h_{b,m} \;\; \forall\, j,b,m\geq 1,
\end{aligned}
\label{eq:13}
\end{equation}
where $c = \{c_{j,b,0}, c_{j,b,1},\ldots,c_{j,b,K}\}$ collects the
coefficients of all $J$ dates and $B$ bands, generalizing the earlier
weighted objective \cite{denaro2026}
%
%
to four independently tunable weights rather than two. The weights serve
distinct roles, which $w_1$ and $w_2$ are required as $w_1$ drives the pairwise
disagreement $E_C$ toward zero and $w_2$ anchors each transform to a common level. Both are set to
$1$. $w_3$ is the spectral-consistency constraint that couples the hinge
coefficients across adjacent bands, so the six bands of a date are
adjusted consistently rather than independently. It is set to $1$, since
lowering it would leave the bands essentially free to move on their own.
$w_4$ penalizes the curvature of the transform across knots, keeping
$T_{j,b}$ smooth rather than kinked where adjacent quantile rows pull in
different . It is set to $0.5$, a mild smoothing that prevents
spurious kinks without preventing the transform from bending where the
data require it.

The key structural property of \eqref{eq:13} is that $T_{j,b}$ is
nonlinear in reflectance but \emph{linear in $c$}, so \eqref{eq:13} is a
bound-constrained weighted least-squares problem, adding the
monotonicity bound \eqref{eq:10} does not change this, since the bound
is itself linear in $c$. Once solved (we use a trust-region
bound-constrained least-squares solver), $T_{j,b}$ is applied to every
pixel of date $j$, band $b$ (not only the DIF subset), producing the
harmonized images used as common input to the nonlinear MGCCA frameworks in Section \ref{sec:nonlinear}. The complete procedure is summarized in Algorithm~\ref{alg:harmonization}.

\begin{algorithm}[t]
\caption{Monotone Piecewise-Linear Harmonization}
\begin{algorithmic}[1]\label{alg:harmonization}
\STATE {\bf Given}
multitemporal images $\mathbf{X}_1,\ldots,\mathbf{X}_J$ ($B$ bands, same stack as Alg.~\ref{alg:deepgcca});
thresholded DIFs from the change-detection flow;
knot count $K$; quantile levels $\{q\}$, $M=41$;
weights $w_1,w_2,w_3,w_4$; minimum slope $\text{min\_slope}>0$.
\STATE
$\mathrm{DIF} \leftarrow$ pixels classified as no-change on \emph{all} $J$ dates
\FOR{$b=1,\ldots,B$}
\STATE
Place knots $t_{b,0}<\cdots<t_{b,K}$ at the quantiles of the pooled DIF population of band $b$ \COMMENT{knots sit where the data are}
\STATE
Build hinge basis $R_1(x;\mathbf{t}_b),\ldots,R_K(x;\mathbf{t}_b)$ \COMMENT{Eq.~\eqref{eq:9}; first/last ramps open-ended}
\ENDFOR
\FOR{$j=1,\ldots,J$, $b=1,\ldots,B$}
\STATE
$Q_{j,b}(q) \leftarrow$ $q$-quantile of $\{X_{j,b,\ell} : \ell \in \mathrm{DIF}\}$, $\forall\,q$ \COMMENT{quantile function of the DIF population}
\ENDFOR
\STATE
Assemble the agreement rows $T_{j,b}(Q_{j,b}(q)) - T_{k,b}(Q_{k,b}(q))$, $j<k$, weighted by $w_1$ \COMMENT{$E_C$, Eq.~\eqref{eq:11}}
\STATE
Assemble the identity-anchor rows $T_{j,b}(Q_{j,b}(q)) - Q_{j,b}(q)$, weighted by $w_2$ \COMMENT{$E_T$, Eq.~\eqref{eq:11}}
\STATE
Assemble the band-smoothness rows $c_{j,b+1,m}-c_{j,b,m}$, weighted by $w_3$ \COMMENT{$E_S$, Eq.~\eqref{eq:12}}
\STATE
Assemble the knot-curvature rows $c_{j,b,m+1}-c_{j,b,m}$, weighted by $w_4$ \COMMENT{$E_K$, Eq.~\eqref{eq:12}}
\STATE
Set lower bounds $c_{j,b,m} \geq \epsilon_b\,h_{b,m}$, $\epsilon_b=\text{min\_slope}$, $h_{b,m}=t_{b,m}-t_{b,m-1}$ \COMMENT{monotonicity as box constraint, Eq.~\eqref{eq:10}}
\STATE
$c^* \leftarrow$ trust-region bound-constrained least-squares solve of $F(c)$ \COMMENT{Eq.~\eqref{eq:13}; linear in $c$ $\Rightarrow$ convex}
\FOR{$j=1,\ldots,J$, $b=1,\ldots,B$}
\STATE
$\mathbf{X}'_{j,b} \leftarrow T_{j,b}(\mathbf{X}_{j,b};\, c^*_{j,b,\cdot})$ applied to \emph{every} pixel \COMMENT{not only the DIF subset}
\ENDFOR
\STATE {\bf Output} harmonized images $\mathbf{X}'_1,\ldots,\mathbf{X}'_J$ aligned to the common level.
\end{algorithmic}
\end{algorithm}

\begin{table*}[htbp]
\centering
\caption{Regions of interest (ROIs), acquisition dates, sensors, spatial resolutions, and environmental descriptions.}
\label{tab:roi_data}
\renewcommand{\arraystretch}{1.2}

\begin{tabular}{c c l l l c l}
\hline
\textbf{No.} & \textbf{ROI} & \textbf{Location} & \textbf{Date} &
\textbf{Sensor} & \textbf{Resolution} & \textbf{Description} \\
\hline
\multirow{4}{*}{1} &
\multirow{4}{*}{
    \includegraphics[width=1.3cm]{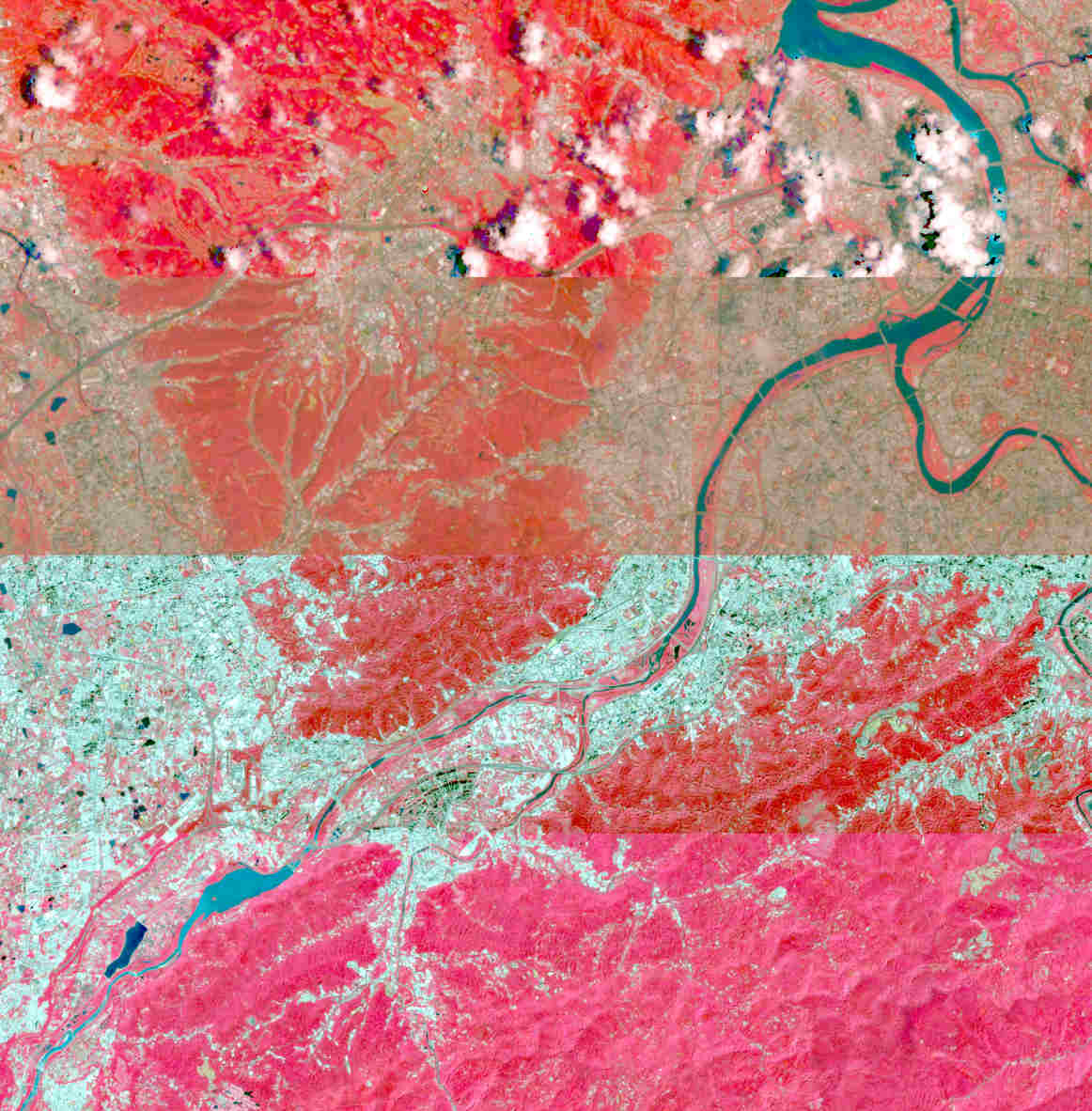}
} &
\multirow{4}{*}{Taipei}
& 2024-08-08 & Landsat 8
& \multirow{4}{*}{10 m}
& \multirow{4}{*}{City and Vegetation} \\
& & & 2024-03-22 & Landsat 9 & & \\
& & & 2024-02-14 & Sentinel-2B & & \\
& & & 2024-09-01 & Sentinel-2B & & \\
\hline
\multirow{4}{*}{2} &
\multirow{4}{*}{
    \includegraphics[width=1.3cm]{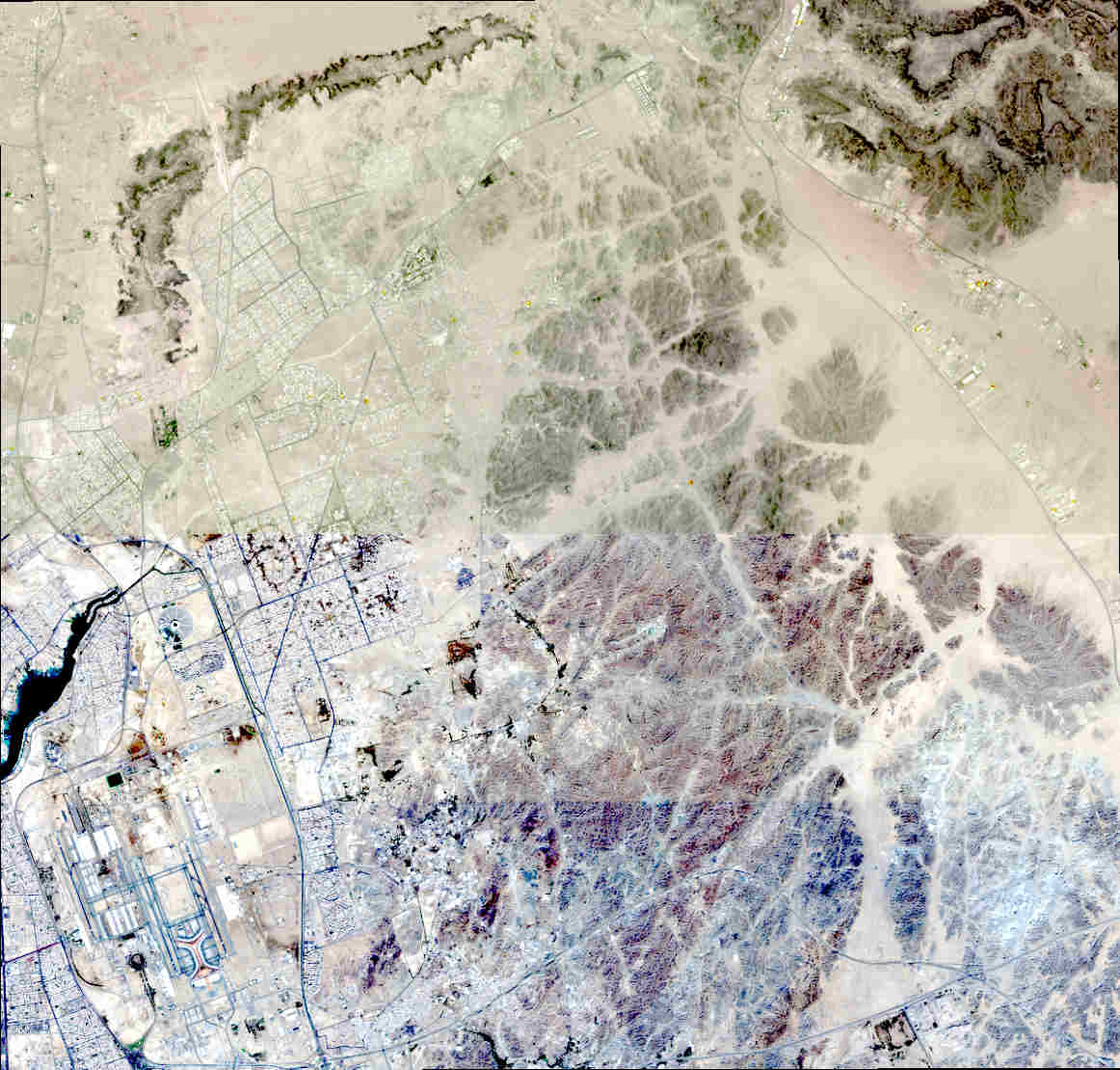}
} &
\multirow{4}{*}{Saudi Arabia}
& 2024-06-20 & Landsat 8
& \multirow{4}{*}{30 m}
& \multirow{4}{*}{City and Desert} \\
& & & 2024-04-25 & Landsat 9 & & \\
& & & 2024-11-06 & Sentinel-2B & & \\
& & & 2024-12-06 & Sentinel-2B & & \\
\hline
\multirow{4}{*}{3} &
\multirow{4}{*}{
    \includegraphics[width=1.3cm]{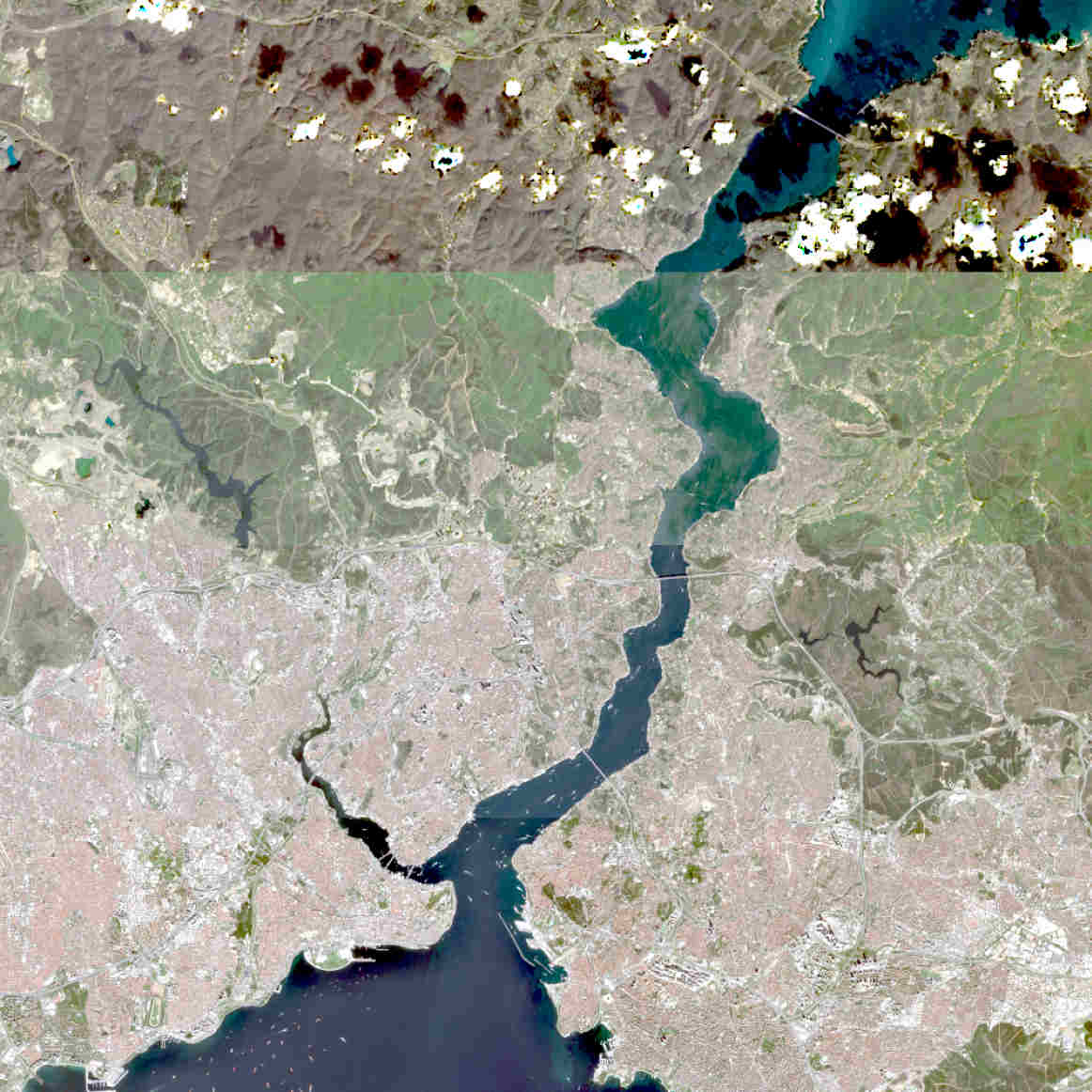}
} &
\multirow{4}{*}{Turkey}
& 2024-06-27 & Landsat 8
& \multirow{4}{*}{30 m}
& \multirow{4}{*}{City and Strait} \\
& & & 2024-04-16 & Landsat 9 & & \\
& & & 2024-03-24 & Sentinel-2A & & \\
& & & 2024-10-05 & Sentinel-2B & & \\
\hline
\multirow{4}{*}{4} &
\multirow{4}{*}{
    \includegraphics[width=1.3cm]{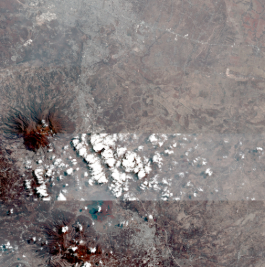}
} &
\multirow{4}{*}{Java}
& 2017-08-22 & Landsat 8
& \multirow{4}{*}{30 m}
& \multirow{4}{*}{City and Forest} \\
& & & 2018-08-25 & Landsat 8 & & \\
& & & 2019-08-28 & Landsat 8 & & \\
& & & 2020-08-30 & Landsat 8 & & \\
\hline
\end{tabular}
\end{table*}

\section{Experimental Results and Discussion}\label{sec: experiment}
\subsection{Datasets}\label{sec:experimental setting}
The dataset used in this study comprises four distinct environmental settings: urban, desert, forest, and coastal. To capture a broad range of geographical, climatic, and environmental conditions, four representative regions of interest (ROIs) were selected as the primary study areas, namely Taipei in Taiwan, Java in Indonesia, Turkey, and Saudi Arabia. These study areas were selected to provide diverse landscape characteristics and environmental conditions, ranging from densely developed urban areas and coastal environments to forested regions and arid desert landscapes. In particular, Taipei represents a predominantly urban environment, Turkey provides an environment characterized by both urban development and a strait, and Saudi Arabia represents an arid environment with urban characteristics, providing a contrasting climatic and landscape context to the study areas in East and Southeast Asia. Java was selected to represent a forest-dominated environment. The diversity of these ROIs enables the dataset to encompass substantial variations in land cover, climate, geographic setting, and levels of urbanization, thereby supporting a more comprehensive evaluation of the proposed approach under different environmental conditions. The acquisition dates, satellite sensors, spatial resolutions, ROI locations, and corresponding environmental descriptions for all study areas are summarized in Table~\ref{tab:roi_data}. 

For all acquisitions, the analysis is restricted to the six spectral bands common to both sensors, blue, green, red, near-infrared (NIR), and the two shortwave-infrared bands (SWIR-1, SWIR-2), whose band numbers, wavelength ranges, and central wavelengths are listed in Table~\ref{tab:bands}. This six-band restriction is deliberate. First, these are the only bands with
near-equivalent counterparts on both Landsat 8/9 OLI and Sentinel-2 MSI: the
Sentinel-2 red-edge bands have no OLI analogue, while the coastal/aerosol and
cirrus bands are either limited to 60~m resolution or dominated by atmospheric
absorption, so a common band set is a prerequisite for cross-sensor DeepMGCCA and MAD statistics. This band set coincides with that of the Harmonized
Landsat--Sentinel-2 (HLS) product and with the majority of harmonization and change-detection studies, making our results directly
comparable to the literature.

\begin{table}[t]
\caption{Spectral bands used in this study. Landsat 8/9 OLI and Sentinel-2 MSI
band numbers, wavelength ranges, and central wavelengths.}
\label{tab:bands}
\renewcommand{\arraystretch}{1.2}
\begin{center}
\setlength{\tabcolsep}{1.8mm}
\begin{tabular}{l c c c c c c}
\hline
\hline
 & \multicolumn{3}{c}{\textbf{Landsat 8/9 OLI}} & \multicolumn{3}{c}{\textbf{Sentinel-2 MSI}} \\
\cmidrule(lr){2-4}\cmidrule(lr){5-7}
\textbf{Band} & \multirow{2}{*}{\textbf{No.}}& \multicolumn{2}{c}{\textbf{Wavelength (nm)}} & \multirow{2}{*}{\textbf{No.}} & \multicolumn{2}{c}{\textbf{Wavelength (nm)}} \\
\cmidrule(lr){3-4}\cmidrule(lr){6-7}
 & & Range & Center & & Range & Center \\
\hline
B1 (Blue)   & B2  & 450--510   & 480  & B2  & 458--523   & 490  \\
B2 (Green)  & B3  & 530--590   & 560  & B3  & 543--578   & 560  \\
B3 (Red)    & B4  & 640--670   & 655  & B4  & 650--680   & 665  \\
B4 (NIR)    & B5  & 850--880   & 865  & B8  & 785--900   & 842  \\
B5 (SWIR-1) & B6  & 1570--1650 & 1610 & B11 & 1565--1655 & 1610 \\
B6 (SWIR-2) & B7  & 2110--2290 & 2200 & B12 & 2100--2280 & 2190 \\
\hline
\hline
\end{tabular}
\end{center}
\end{table}

\subsection{Experimental result and comparison}\label{sec:ablation}
We evaluate the proposed DeepMGCCA, whose three-block residual shared encoder
is followed by a nonlinear per-date head output. The encoder is trained for ten epochs under the iterative
reweighting scheme, after which the consensus-aligned variates $\mathbf{U}_j$
yield a global DIF set. The harmonization
transform is fitted on the quantile population of this DIF set and then
applied to every pixel of the scene.

\begin{figure*}[t]
\centering
\includegraphics[width=1\linewidth]{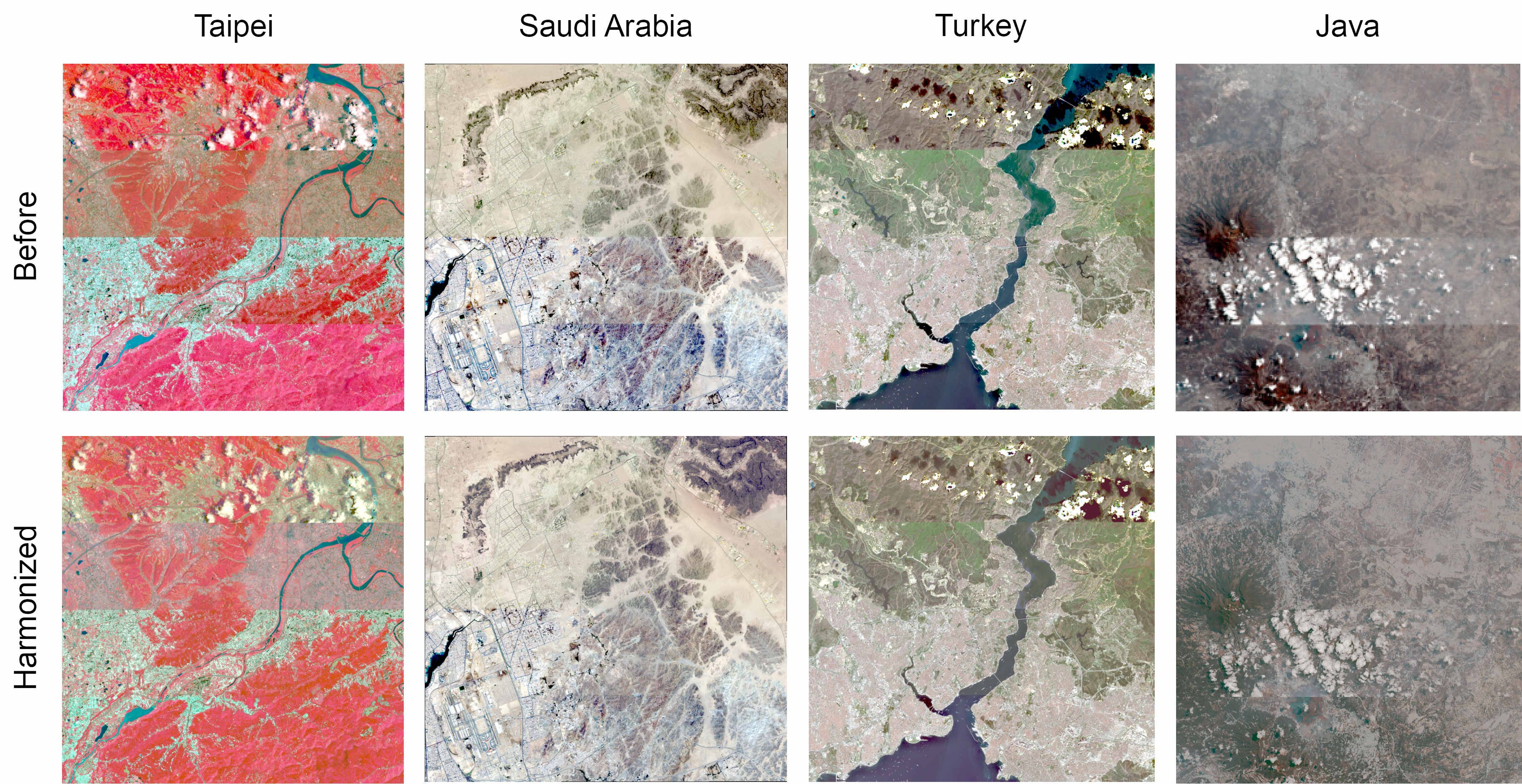}
\caption{Qualitative harmonization results for the four ROIs of
Table~\ref{tab:roi_data}. For each ROI, the four multitemporal acquisitions
are mosaicked into false-/true-color composites: (\textit{top}) the original
surface-reflectance images, in which interdate brightness and color seams are
clearly visible across acquisition dates, and (\textit{bottom}) the harmonized
images produced by the proposed transform, in which the interdate drift is
removed while genuine spatial detail is preserved.}
\label{fig:mosaic}
\end{figure*}

\begin{figure*}[t]
\centering
\includegraphics[width=0.76\linewidth, angle=-90]{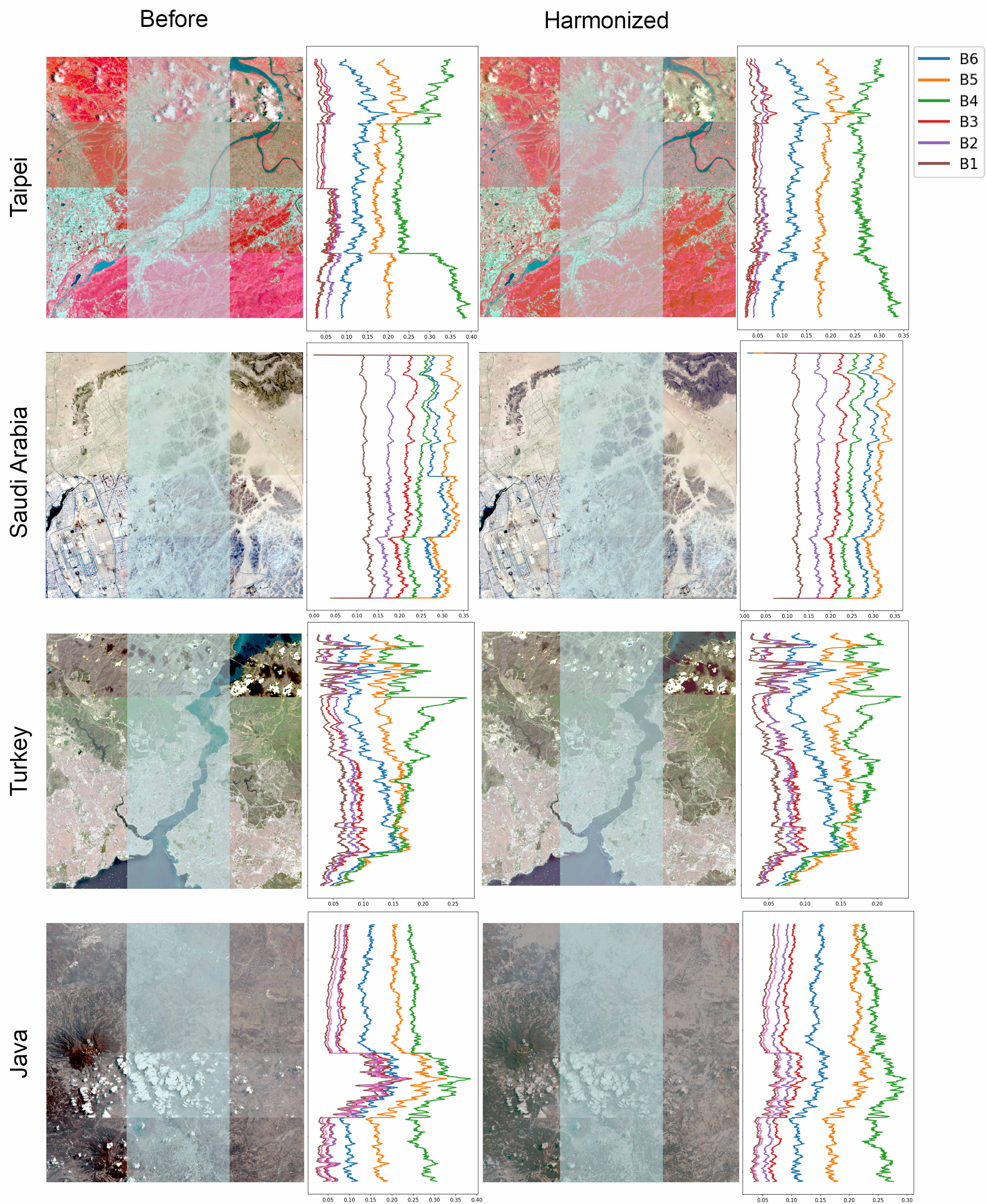}
\caption{Row-wise mean reflectance profiles over the central column window (highlighted as the translucent band in the composites), computed for the four ROIs of Table~\ref{tab:roi_data} from the four multitemporal acquisitions of each ROI (\textit{left}: original and \textit{right}: harmonized by the proposed transform). For each scene, the per-row mean reflectance (horizontal axis) is plotted against the image row (vertical axis, top to bottom), with one curve per acquisition date and spectral band. Before harmonization, the four date-profiles appear as vertically shifted copies of one another, a direct visualization of the interdate reflectance inconsistency (additive and multiplicative bias) on invariant terrain. After harmonization, the profiles of the four dates seamed, indicating that the interdate drift is removed without suppressing genuine spatial or temporal detail.}
\label{fig:profile}
\end{figure*}

Fig.~\ref{fig:mosaic} shows the mosaicked false-/true-color composites of the four multitemporal acquisitions for each ROI of Table~\ref{tab:roi_data}, before and after harmonization. In the original composites, the boundaries between the acquisition dates are marked by visible brightness and color seams. Each rectangle of the mosaic carries its own temporal level, so terrain and water bodies that cross a date boundary change appearance abruptly even though the surface itself did not change. After applying the proposed transform fitted on the DeepMGCCA DIFs, these seams are visibly removed and the mosaic becomes temporally continuous, the same ground features keep the same color and brightness across the date boundaries, while fine spatial texture such as roads, buildings, and field boundaries remains sharp and is not flattened by the correction. Fig.~\ref{fig:profile} depicts the corresponding row-wise mean reflectance profiles over the central column window of each scene, which make the interdate drift explicit as vertical shifts of the per-band curves at the rectangle boundaries. Before harmonization, the band profiles are clearly shifted from one date segment to the next, consistent with residual additive and multiplicative bias between acquisition dates. After harmonization, this shift is strongly reduced and the profiles of the four dates overlap along the whole row axis, indicating that the interdate drift has been removed on invariant terrain without suppressing genuine spatial or temporal variation. The pronounced decrease of the profile values toward the bottom rows of the Turkey scene is not an artifact of the method. The lower part of this ROI is dominated by the strait, whose low reflectance pulls the row-wise mean down, and the same sea-dominated segment also explains why the four dates were the least consistent there before harmonization. Taken together, these qualitative results indicate that the DIF-guided transform removes interdate bias without flattening spatial texture or altering genuinely changed surfaces. The quantitative evaluation of the harmonization variants is deferred to Section~\ref{sec:performance}.

\subsection{Performance evaluation}\label{sec:performance}

\begin{table*}[t]
\caption{Per-ROI quantitative comparison of harmonization variants for the proposed DeepMGCCA. Values are means over all date pairs and spectral bands. RMSE, SAM$_{h}$, and SAM$_{sc}$, lower indicates better result.  PSNR, SSIM, and Pearson's $\rho$,  higher indicates better result. SAM$_{h}$ is the mean spectral angle between the harmonized dates over the DIF population. SAM$_{sc}$ is the mean spectral angle between each image and its own corrected version, undefined for the unprocessed images (---). Orig. is unprocessed. RA is reference-affine \cite{DU2002123}.  SSA is sequential stepwise affine \cite{lin2019}. GSC is globally spectral constrained \cite{denaro2026}. Proposed is the proposed monotone piecewise-linear transform. Best value per column per ROI in bold. in the SAM$_{sc}$ column both GSC and Proposed are highlighted, as the two operate on a deliberate trade-off between spectral self-preservation and between-date agreement (see text). All variants are scored on the same DIF mask.}
\label{tab:quanti}
\centering
\renewcommand{\arraystretch}{1.2}
\setlength{\tabcolsep}{2.2mm}
\small
\begin{tabular}{c l c c c c c c}
\hline\hline
ROI & Variant & RMSE$\downarrow$ & PSNR$\uparrow$ & SSIM$\uparrow$ & SAM$_{h}$$\downarrow$ & SAM$_{sc}$$\downarrow$ & $\rho\uparrow$ \\
\hline
\multirow{5}{*}{\emph{Taipei}}
 & Orig.    & 0.0359 & 31.91 & 0.7568 & 5.445 & --- & {\bf 0.7416} \\
 & RA\cite{DU2002123}   & 0.0217 & 35.33 & 0.7894 & {\bf 2.512} & 4.872 & 0.7407 \\
 & SSA \cite{lin2019} & 0.0222 & 35.26 & 0.7882 & 2.522 & 4.926 & 0.7407 \\
 & GSC\cite{denaro2026}    & 0.0238 & 34.84 & 0.8065 & 3.630 & {\bf 2.262} & 0.7409 \\
 & Proposed    & {\bf 0.0210} & {\bf 35.81} & {\bf 0.8184} & {\bf 2.907} & {\bf 2.837} & 0.7106 \\
\hline
\multirow{5}{*}{\emph{Saudi Arabia}}
 & Orig.    & 0.0237 & 34.17 & 0.9230 & 3.066 & --- & 0.9499 \\
 & RA\cite{DU2002123}   & 0.0164 & 36.51 & 0.9588 & 1.380 & 2.253 & 0.9552 \\
 & SSA\cite{lin2019} & 0.0160 & 36.84 & 0.9593 & 1.442 & 2.139 & 0.9552 \\
 & GSC\cite{denaro2026}    & 0.0148 & 37.48 & 0.9564 & 1.562 & {\bf 1.287} & 0.9552 \\
 & Proposed    & {\bf 0.0138} & {\bf 38.21} & {\bf 0.9618} & {\bf 1.292} & {\bf 1.530} & {\bf 0.9565} \\
\hline
\multirow{5}{*}{\emph{Turkey}}
 & Orig.    & 0.0354 & 30.57 & 0.7694 & 7.535 & --- & 0.7304 \\
 & RA\cite{DU2002123}   & 0.0328 & 31.59 & 0.7768 & 6.764 & 5.240 & 0.7272 \\
 & SSA\cite{lin2019} & 0.0292 & 32.43 & 0.7894 & 6.742 & 4.882 & 0.7273 \\
 & GSC\cite{denaro2026}    & 0.0296 & 31.74 & 0.7741 & 6.607 & {\bf 2.907} & 0.7283 \\
 & Proposed    & {\bf 0.0271} & {\bf 32.71} & {\bf 0.7948} & {\bf 5.913} & {\bf 3.940} & {\bf 0.7346} \\
\hline
\multirow{5}{*}{\emph{Java}}
 & Orig.    & 0.0230 & 33.74 & 0.8747 & 5.141 & --- & 0.8501 \\
 & RA\cite{DU2002123}   & 0.0210 & 34.22 & 0.8867 & 4.126 & 4.068 & 0.8499 \\
 & SSA \cite{lin2019} & 0.0220 & 33.72 & 0.8822 & 4.489 & 3.932 & 0.8495 \\
 & GSC\cite{denaro2026}    & 0.0198 & 34.99 & 0.8886 & 4.038 & {\bf 1.770} & 0.8476 \\
 & Proposed    & {\bf 0.0181} & {\bf 35.83} & {\bf 0.9021} & {\bf 3.719} & {\bf 2.327} & {\bf 0.8597} \\
\hline\hline
\end{tabular}
\end{table*}

In this evaluation, six quantities are reported, each capturing a different
aspect of the harmonization. RMSE and PSNR measure the radiometric
agreement between the dates over the DIF population, in reflectance units
and in decibels, respectively. SSIM measures the agreement of local
spatial structure. Pearson's $\rho$ measures the linear correlation of the
per-band scatter between dates. The spectral angle is reported in two
directions with different purposes. SAM$_{sc}$ \cite{denaro2026}, computed between each image and its own corrected version, $\mathbf{X}'_j$ versus $\mathbf{X}_j$, measures how much spectral distortion the transform itself introduces, and therefore judges the self-consistency of each correction. Motivated by the metric \cite{denaro2026}, in this paper, we add metric evaluation as SAM$_{h}$, computed between the
harmonized dates, which is $\mathbf{X}'_j$ versus $\mathbf{X}'_k$, $\text{SAM}_{h}=\frac{1}{J}\sum{^{J}_{j=1,k\ne j}\text{SAM}(\mathbf{X}_j',\mathbf{X}_k')}$, measures how
well the correction has aligned the spectral signatures across time, and
is therefore the primary objective of harmonization. A successful harmonization should minimize while keeping
SAM$_{sc}$ small.

Table~\ref{tab:quanti} reports the per-ROI comparison, all variants fitted
on the same DIF population and scored on identical PIF masks. Averaging
the four ROI means, the proposed transform lowers the RMSE from $0.0295$
to $0.0200$ ($-32\%$), lowers the between-date spectral angle SAM$_{h}$
from $5.30^\circ$ to $3.46^\circ$ ($-35\%$), and raises the PSNR from
$32.60$ to $35.64$~dB and the SSIM from $0.8310$ to $0.8693$. It attains
the best RMSE, PSNR, and SSIM in all four ROIs, the best SAM$_{h}$ in
Saudi Arabia, Turkey, and Java, and the best Pearson's $\rho$ in the three
30~m. In the SAM$_{sc}$ column it is highlighted together with globally spectral constrained (GSC) \cite{denaro2026},
for the reason explained next.

The comparison between GSC and the proposed transform is a systematic
trade-off, identical in direction in every ROI and therefore not
attributable to chance. GSC attains the lowest self-distortion (Saudi
Arabia $1.29^\circ$, Turkey $2.91^\circ$, Java $1.77^\circ$, Taipei
$2.26^\circ$ and their averaged over the ROIs is $2.06^\circ$), while the proposed
transform attains the lower between harmonized-date angle (Saudi $1.29^\circ$
vs.\ $1.56^\circ$, Turkey $5.91^\circ$ vs.\ $6.61^\circ$, Java
$3.72^\circ$ vs.\ $4.04^\circ$, Taipei $2.91^\circ$ vs.\ $3.63^\circ$ and their 
averaged is $3.46^\circ$ vs.\ $3.96^\circ$) together with the best RMSE,
PSNR, and SSIM. Crucially, the higher SAM$_{sc}$ of the
proposed transform does not mean that it fails to preserve the spectral
signature. Fig.~\ref{fig:profile} shows that the per-band reflectance
profiles remain aligned before and after harmonization, i.e., each band
keeps its relative position across the spectrum, and the transform is by
construction monotone with a strictly positive minimum slope, so no two
reflectance levels are ever collapsed and the spectral ordering is
preserved exactly. The $0.60^\circ$ average difference from GSC is
therefore not misalignment but the capacity. The piecewise-linear transformation
simply has more freedom under the constraints, and it spends that freedom
on removing more interdate bias, while staying within the tolerance allowed by the monotonicity and smoothness constraints. The two methods are best read as two points on a Pareto front, GSC at the
spectral-preservation end, the proposed transform at the high-correction
end, and both are defensible operating points, which is why both are
highlighted in Table~\ref{tab:quanti}. Taipei is the one scene where the
between-date angle is minimized by the reference-based fits (reference-affine (RA) \cite{DU2002123}
$2.51^\circ$, sequential stepwise affine (SSA) \cite{lin2019}  $2.52^\circ$). It is also the only scene where the
proposed transform's Pearson's $\rho$ decreases ($0.7416 \to 0.7106$)
while all other methods remain within $0.0009$ of the original. Both effects are consistent with Taipei's 10~m pansharpened data, whose high-frequency per-band variations are re-weighted by any nonlinear warp
in ways a linear statistic cannot credit. Pearson's $\rho$ measures only
linear association, so a transform that straightens a mildly curved
relationship can nonetheless lower it, whereas a rank-based correlation,
which any strictly monotone per-band transform preserves by construction,
would show no such drop.

The reference-based variants, by contrast, exhibit the failure mode that
the global design of GSC and of the proposed transform avoids. Both RA and
SSA align every date to a single reference, so fit quality decays with
radiometric distance from that reference, and SSA compounds this by
chaining each date to the already-normalized previous one, so any error in
one step is inherited by all later dates. The data show the predicted
signature: in Turkey, the scene with the widest interdate range, RA
produces the largest self-distortion of any method ($5.24^\circ$), and in
Java SSA's PSNR ($33.72$~dB) falls below that of the unprocessed images
($33.74$~dB): the accumulated chain error exceeds the bias it removes.
Finding the common level globally, through all-pairs agreement rather
than a reference selection, is what keeps both GSC and the proposed
transform ahead of the reference-based variants on nearly every metric in
every ROI.

\subsection{Discussion}\label{sec:discussion}

\subsubsection{DeepMGCCA architecture and training}
The central design choice is to separate the shared nonlinear feature
extractor from the date-specific projection. One encoder $f_{\theta}$ is
applied to every date, and a head $h_{\phi_j}$ produces the
head outputs $\mathbf{Z}_{j}=h_{\phi_j}(f_{\theta}(\mathbf{X}_{j}))$ of
\eqref{eq:canonical-variates}. What is common to all dates is learned
once, in $\theta$. The heads learn only the residual date-specific
behavior that the MAD statistic must expose, and they generalize the
per-date projections $\mathbf{A}_{j}$ of \eqref{eq:maxvar}, keeping the
model interpretable as canonical analysis on learned features. Granting
each date a full encoder would let the optimizer raise agreement by
warping changed pixels onto their counterparts, so every head convolution
is constrained to $1\times1$ with no spatial reach, agreement must be
earned from the shared features themselves. The learning signal is the
correlation itself, so the pipeline is fully unsupervised and, as recent
evidence supports, a simple convolutional encoder suffices for this task.

This freedom still carries a collapse risk. Nothing in the correlation
objective checks that the features describe the input, and an encoder
that maps every pixel to almost the same output maximizes agreement
while destroying the change information. A small per-date decoder,
trained to reconstruct $\mathbf{X}_{j}$ from $\mathbf{Z}_{j}$
(\eqref{eq:recon}), is what prevents this. If decoder coefficient is set to 0,
reconstruction fails, so the only way to keep both terms small is to
keep $\mathbf{Z}_{j}$ informative in the latent space. The weight $\lambda_{\mathrm{recon}}$
sets how strictly this is enforced. At $0$ nothing stops collapse, at
very large values the model becomes an autoencoder, and the implemented
value $\lambda_{\mathrm{recon}}=1$ gives the two terms comparable weight,
with the final DIF quality insensitive over $[0.5,2]$. The reconstruction
curve is logged separately so a collapsing run is visible in the log
rather than discovered afterwards.

\subsubsection{Objective, weighting, and reliability}
Among the GCCA criteria, maximum-variance \eqref{eq:maxvar} is solved in closed
form at each step through an eigenproblem on a $(Jr\times Jr)$ matrix, so
only the network parameters carry gradients and the eigendecomposition is
never differentiated through. Its consensus $\mathbf{G}$ fixes the
cross-date component correspondence during training, making the
difference variates of \eqref{eq:mad-deep} meaningful by construction,
and the constraint \eqref{eq:weighted-constraint} is enforced exactly by
weighted center-scaling. The weights themselves are the IR-MAD robustness
mechanism that seeded at $t=0$ by a temporal and a correlation-scaled
spectral-angle factor \eqref{eq:tfactor}--\eqref{eq:sam-init} and updated
from $t\geq1$ by the NMAD rule \eqref{eq:inv-nmad}, they define the
training covariances, the constraint, and the final detection statistic
at once. Because high agreement can mean either genuine convergence or
collapse, two diagnostics arbitrate reliability. The ratio of the
objective under uniform versus robust weights, and the ratio of raw-band
cross-date RMSE between the least- and most-weighted deciles, which
verifies on the original reflectance that the trusted pixels really are
the invariant ones.

This closed-form solve is also what keeps the deep extension scalable
where the kernel extension \eqref{eq:kgcca} is not. The eigenproblem is independent of the pixel count and only the per-date covariances grow,
at $O(Nr^{2})$, so million-pixel scenes are processed at modest cost
while kernel GCCA's $O(n^{2})$ memory and $O(n^{3})$ time quickly become
prohibitive.

\subsubsection{Thresholding and harmonization}

The change score of a pixel is the sum of the squared, standardized
canonical differences, aggregated over the $r$ components and the
connected date pairs \eqref{eq:mad-deep}. Under the null hypothesis that
the pixel is invariant, each standardized difference is approximately a
standard normal variate, so the score itself follows a chi-square
distribution whose degrees of freedom equal the number of aggregated
terms \eqref{eq:nmad}. The threshold is therefore applied to this
statistic, not to reflectance. The chi-square survival function converts
each pixel's score into the tail probability of the no-change hypothesis,
$\Pr(\chi^{2}_{d} \geq \mathrm{nmad}_{\ell})$ \eqref{eq:inv-nmad}, and
the DIF set is built from the pixels whose score falls below the same
right-tail cutoff of the null distribution. This makes the threshold
\emph{dynamic} for a precise statistical reason. The cutoff is a fixed
property of the chi-square tail, but the mapping from reflectance to
score differs from scene to scene, so the same decision rule
automatically admits a wider no-change region in a noisy scene and a
narrower one in a stable scene. Every scene is judged against the null
distribution of its own data, which is why scenes with very different
bias levels still contribute comparable invariant populations and all
variants can be scored on same-sized DIF sets. Ties at the cutoff, which
the chi-square map can produce in large plateaus, are broken by a seeded
random draw rather than by image order, keeping the selection
reproducible. 

The harmonization stage then fits the monotone piecewise-linear transform
of \eqref{eq:9}--\eqref{eq:13}. The quantile-placed knots, monotonicity as a
box constraint with a positive minimum slope, and a convex four-term
objective whose identity reference finds the common level globally, without
a reference date. This global design avoids the failure mode of the
reference-based affine variants \cite{DU2002123}, whose accuracy decays with distance from
the chosen reference and, in the sequential variant \cite{lin2019}, compounds along the
date chain until the correction becomes worse than none. Against the
globally spectral-constrained variant, the proposed transform accepts
slightly higher self-distortion SAM$_{sc}$ in exchange for a lower
between-date angle SAM$_{h}$ and the best RMSE, PSNR, and SSIM in every
ROI. Since the profiles of Fig.~\ref{fig:profile} remain aligned and
monotonicity preserves the spectral ordering exactly, the extra freedom
is spent removing bias, not misaligning bands. The two methods are two
defensible points on one Pareto front, which is why both are highlighted
in Table~\ref{tab:quanti}, and because $K{=}1$ reduces the transformation to the
affine map, the transform stays accurate even where the relationship is
nearly linear, engaging its extra knots only where the data bend.

\section{Conclusions and Future Works}\label{sec:conclusion}

This study introduced DeepMGCCA, a deep multiview invariant-feature learning framework for unsupervised multitemporal surface reflectance harmonization that unifies nonlinear invariant-feature extraction, robust iterative reweighting, and piecewise constrained harmonization. Replacing the fixed linear projection of classical GCCA with a shared convolutional encoder followed by date-specific nonlinear head outputs, the framework learns DIFs directly from pixel-weighted canonical variates, with the unit-variance constraint enforced by construction through weighted center-scaling and a reconstruction regularizer preventing the free nonlinear reparameterizations of the heads from collapsing genuine temporal variability.
Building on the extracted DIFs, a piecewise constrained harmonization uses monotone piecewise-linear transform, fitted on the DIF quantile population and estimated as a bound-constrained least-squares problem, removes interdate bias while preserving temporal-spectral consistency by construction.  
Evaluated on four ROIs spanning urban, desert, coastal, and forest environments, the proposed transform attained the best RMSE, PSNR, and SSIM in every ROI and the best between-date spectral angle and Pearson's $\rho$ in three of the four, lowering the ROI-mean RMSE by $32\%$ and the between-date spectral angle by $35\%$. These gains come with a consistent trade-off against the GSC and the proposed method. The proposed transform accepts slightly higher self-distortion in exchange for a larger removal of interdate bias, and the per-band profiles confirm that the spectra still remain aligned before and after harmonization. The two methods are therefore best read as two defensible points on a Pareto front. These results demonstrate that deep invariant-feature learning and convex harmonization can be combined into a single unsupervised pipeline that is expressive enough to model nonlinear interdate structure and constrained enough their spectral consistency.
Future work will pursue CODE (Convex/Deep Learning \cite{CODE}) harmonization that jointly optimizing the convex transform and the deep invariant-feature learner in a single objective, so that the harmonization criterion shapes the very features from which the DIFs are drawn.


\bibliography{ref}

@article{HyperQUEEN,
  title={{HyperQUEEN:} {H}yperspectral quantum deep network for image restoration},
  author={Lin, Chia-Hsiang and Chen, You-Yao},
  journal={IEEE Trans. Geosci. Remote Sens.},
  volume={61},
  pages={1--20},
  year={May 2023},
  publisher={IEEE}
}

@INPROCEEDINGS{Daudt2018,
  author={Caye Daudt, Rodrigo and Le Saux, Bertr and Boulch, Alexandre},
  booktitle={Proc. IEEE International Conference on Image Processing (ICIP)}, 
  title={Fully Convolutional Siamese Networks for Change Detection}, 
  year={07-10 Oct. 2018},
  address={Athens, Greece},
  volume={},
  number={},
  pages={4063-4067},
  doi={10.1109/ICIP.2018.8451652}}

@Article{Chen2020,
AUTHOR = {Chen, Hao and Shi, Zhenwei},
TITLE = {A Spatial-Temporal Attention-Based Method and a New Dataset for Remote Sensing Image Change Detection},
JOURNAL = {Remote Sensing},
VOLUME = {12},
YEAR = {Apr. 2020},
NUMBER = {10},
ARTICLE-NUMBER = {1662},
URL = {https://www.mdpi.com/2072-4292/12/10/1662},
ISSN = {2072-4292},
DOI = {10.3390/rs12101662}
}

@ARTICLE{Khelifi2020,
  author={Khelifi, Lazhar and Mignotte, Max},
  journal={IEEE Access}, 
  title={Deep Learning for Change Detection in Remote Sensing Images: Comprehensive Review and Meta-Analysis}, 
  year={Jul. 2020},
  volume={8},
  number={},
  pages={126385-126400},
  doi={10.1109/ACCESS.2020.3008036}}

@Article{Shi2020,
AUTHOR = {Shi, Wenzhong and Zhang, Min and Zhang, Rui and Chen, Shanxiong and Zhan, Zhao},
TITLE = {Change Detection Based on Artificial Intelligence: State-of-the-Art and Challenges},
JOURNAL = {Remote Sensing},
VOLUME = {12},
YEAR = {Apr. 2020},
NUMBER = {10},
ARTICLE-NUMBER = {1688},
URL = {https://www.mdpi.com/2072-4292/12/10/1688},
ISSN = {2072-4292},
DOI = {10.3390/rs12101688}
}

@article{IWDQUEEN,
  title={Quantum-Driven Multihead Inland Waterbody Detection With Transformer-Encoded {CYGNSS delay-Doppler} Map Data},
  author={Lin, Chia-Hsiang and Lin, Jhao-Ting and Chiu, Po-Ying and Chen, Shih-Ping and Lin, Charles CH},
  journal={IEEE Trans. Geosci. Remote Sens.},
  volume={63},
  pages={1--21},
  year={Nov. 2025},
  publisher={IEEE}
}

@article{tang2025quantum,
  title={A quantum-empowered {SPEI} drought forecasting algorithm using spatially aware mamba network},
  author={Tang, Po-Wei and Lin, Chia-Hsiang and Huang, Jian-Kai and Huete, Alfredo R},
  journal={IEEE Trans. Geosci. Remote Sens.},
  volume={63},
  pages={1--18},
  year={Jun. 2025},
  publisher={IEEE}
}

@ARTICLE{HyperLUCID,
  author={Lin, Chia-Hsiang and Hsu, Shih-Min and Liang, Ching-Yun and Chanussot, Jocelyn and Chen, Jhih-Yan},
  journal={IEEE Transactions on Image Processing}, 
  title={Hyperspectral Calibration Detection: A Novel Concept for Change Detection With Unsupervised Incremental Safe Pseudo-Labeling Implementation}, 
  year={Aug. 2026},
  volume={35},
  number={},
  pages={8835-8850},
  doi={10.1109/TIP.2026.3720508}}

@ARTICLE{Xu2023,
  author={Xu, Hanzeyu and Zhou, Yuyu and Wei, Yuchun and Liu, Chong and Li, Xiao and Chen, Wei},
  journal={IEEE J. Sel. Topics Appl. Earth Observ. Remote Sens. Observations and Remote Sensing}, 
  title={A Relative Radiometric Normalization Method for Enhancing Radiometric Consistency of Landsat Time-Series Imageries}, 
  year={Jun. 2023},
  volume={16},
  number={},
  pages={5797-5812},
  doi={10.1109/JSTARS.2023.3288973}}

@article{SCHOTT19881,
title = {Radiometric scene normalization using pseudoinvariant features},
journal = {Remote Sensing of Environment},
volume = {26},
number = {1},
pages = {1-16},
year = {Oct. 1988},
issn = {0034-4257},
doi = {https://doi.org/10.1016/0034-4257(88)90116-2},
url = {https://www.sciencedirect.com/science/article/pii/0034425788901162},
author = {John R. Schott and Carl Salvaggio and William J. Volchok}
}

@article{HALL199111,
title = {Radiometric rectification: Toward a common radiometric response among multidate, multisensor images},
journal = {Remote Sensing of Environment},
volume = {35},
number = {1},
pages = {11-27},
year = {Jan. 1991},
issn = {0034-4257},
doi = {https://doi.org/10.1016/0034-4257(91)90062-B},
url = {https://www.sciencedirect.com/science/article/pii/003442579190062B},
author = {F.G. Hall and D.E. Strebel and J.E. Nickeson and S.J. Goetz}
}

@ARTICLE{Liu202000,
  author={Liu, Kunbo and Ke, Tao and Tao, Pengjie and He, Jianan and Xi, Ke and Yang, Kaijun},
  journal={IEEE J. Sel. Topics Appl. Earth Observ. Remote Sens. Observations and Remote Sensing}, 
  title={Robust Radiometric Normalization of Multitemporal Satellite Images Via Block Adjustment Without Master Images}, 
  year={Oct. 2020},
  volume={13},
  number={},
  pages={6029-6043},
  doi={10.1109/JSTARS.2020.3028062}}

@article{CANTY2004441,
title = {Automatic radiometric normalization of multitemporal satellite imagery},
journal = {Remote Sensing of Environment},
volume = {91},
number = {3},
pages = {441-451},
year = {Jun. 2004},
issn = {0034-4257},
doi = {https://doi.org/10.1016/j.rse.2003.10.024},
url = {https://www.sciencedirect.com/science/article/pii/S0034425704001208},
author = {Morton J Canty and Allan A Nielsen and Michael Schmidt}
}

@article{CANTY20081025,
title = {Automatic radiometric normalization of multitemporal satellite imagery with the iteratively re-weighted MAD transformation},
journal = {Remote Sensing of Environment},
volume = {112},
number = {3},
pages = {1025-1036},
year = {Mar. 2008},
issn = {0034-4257},
doi = {https://doi.org/10.1016/j.rse.2007.07.013},
url = {https://www.sciencedirect.com/science/article/pii/S0034425707003495},
author = {Morton J. Canty and Allan A. Nielsen}
}

@article{TENENHAUS2015114,
title = {Kernel Generalized Canonical Correlation Analysis},
journal = {Computational Statistics \& Data Analysis},
volume = {90},
pages = {114-131},
year = {2015},
issn = {0167-9473},
doi = {https://doi.org/10.1016/j.csda.2015.04.004},
url = {https://www.sciencedirect.com/science/article/pii/S0167947315001012},
author = {Arthur Tenenhaus and Cathy Philippe and Vincent Frouin}
}

@article{NIELSEN19981,
title = {Multivariate Alteration Detection {(MAD)} and {MAF} Postprocessing in Multispectral, Bitemporal Image Data: New Approaches to Change Detection Studies},
journal = {Remote Sensing of Environment},
volume = {64},
number = {1},
pages = {1-19},
year = {Apr. 1998},
issn = {0034-4257},
doi = {https://doi.org/10.1016/S0034-4257(97)00162-4},
url = {https://www.sciencedirect.com/science/article/pii/S0034425797001624},
author = {Allan A. Nielsen and Knut Conradsen and James J. Simpson}
}

@incollection{hotelling1992relations,
  title={Relations {B}etween {T}wo {S}ets of {V}ariates},
  author={Hotelling, Harold},
  booktitle={Breakthroughs in Statistics: Methodology and Distribution},
  pages={162--190},
  year={1992},
  publisher={Springer}
}

@ARTICLE{NielsenAasbjerg2007,
  author={Nielsen, Allan Aasbjerg},
  journal={IEEE Transactions on Image Processing}, 
  title={The Regularized Iteratively Reweighted MAD Method for Change Detection in Multi- and Hyperspectral Data}, 
  year={Feb. 2007},
  volume={16},
  number={2},
  pages={463-478},
  doi={10.1109/TIP.2006.888195}}

@ARTICLE{BruzzonePrieto2000,
  author={Bruzzone, L. and Prieto, D.F.},
  journal={IEEE Trans. Geosci. Remote Sens.}, 
  title={Automatic analysis of the difference image for unsupervised change detection}, 
  year={May 2000},
  volume={38},
  number={3},
  pages={1171-1182},
  doi={10.1109/36.843009}}

@ARTICLE{CelikTurgay2009,
  author={Celik, Turgay},
  journal={IEEE Geoscience and Remote Sensing Letters}, 
  title={Unsupervised Change Detection in Satellite Images Using Principal Component Analysis and $k$-Means Clustering}, 
  year={Aug. 2009},
  volume={6},
  number={4},
  pages={772-776},
  doi={10.1109/LGRS.2009.2025059}}

@article{Horst_1961, title={Relations among m Sets of Measures}, volume={26}, DOI={10.1007/BF02289710}, number={2}, journal={Psychometrika}, author={Horst, Paul}, year={1961}, pages={129–149}}

@ARTICLE{lin2019,
  author={Lin, Bo-Yi and Wang, Zhi-Jia and Syariz, Muhammad Aldila and Denaro, Lino Garda and Lin, Chao-Hung},
  journal={IEEE Geoscience and Remote Sensing Letters}, 
  title={Pseudoinvariant Feature Selection Using Multitemporal MAD for Optical Satellite Images}, 
  year={Mar. 2019},
  volume={16},
  number={9},
  pages={1353-1357},
  doi={10.1109/LGRS.2019.2899969}}

@ARTICLE{denaro2026,
  author={Denaro, Lino Garda and Lin, Chao-Hung},
  journal={IEEE J. Sel. Topics Appl. Earth Observ. Remote Sens. Observations and Remote Sensing}, 
  title={Constrained Harmonization of Optical Satellite Imagery for Surface Reflectance Bias Removal}, 
  year={May 2026},
  volume={19},
  number={},
  pages={16779-16792},
  doi={10.1109/JSTARS.2026.3689900}}

@ARTICLE{Denaro2020,
  author={Denaro, Lino Garda and Lin, Chao-Hung},
  journal={IEEE J. Sel. Topics Appl. Earth Observ. Remote Sens. Observations and Remote Sensing}, 
  title={Hybrid Canonical Correlation Analysis and Regression for Radiometric Normalization of Cross-Sensor Satellite Imagery}, 
  year={Feb. 2020},
  volume={13},
  number={},
  pages={976-986},
  doi={10.1109/JSTARS.2020.2971857}}

@article{SYARIZ201956,
title = {{S}pectral-consistent relative radiometric normalization for multitemporal {L}andsat-8 imagery},
journal = {ISPRS Journal of Photogrammetry and Remote Sensing},
volume = {147},
pages = {56-64},
year = {Nov. 2019},
issn = {0924-2716},
doi = {https://doi.org/10.1016/j.isprsjprs.2018.11.007},
url = {https://www.sciencedirect.com/science/article/pii/S0924271618303034},
author = {Muhammad Aldila Syariz and Bo-Yi Lin and Lino Garda Denaro and Lalu Muhamad Jaelani and Manh {Van Nguyen} and Chao-Hung Lin}
}

@article{Tenenhaus_Tenenhaus_2011, title={Regularized Generalized Canonical Correlation Analysis}, volume={76}, DOI={10.1007/s11336-011-9206-8}, number={2}, journal={Psychometrika}, author={Tenenhaus, Arthur and Tenenhaus, Michel}, year={2011}, pages={257–284}}

@ARTICLE{Nielsen2002,
  author={Nielsen, A.A.},
  journal={IEEE Transactions on Image Processing}, 
  title={Multiset canonical correlations analysis and multispectral, truly multitemporal remote sensing data}, 
  year={Mar. 2002},
  volume={11},
  number={3},
  pages={293-305},
  doi={10.1109/83.988962}}

@inproceedings{nielsen2010linear,
  title={Linear and kernel methods for multi-and hypervariate change detection},
  author={Nielsen, Allan A and Canty, Morton J},
  booktitle={Image and Signal Processing for Remote Sensing XVI},
  volume={7830},
  pages={233--239},
  year={Oct. 2010},
  organization={SPIE}
}

@INPROCEEDINGS{DenaroC2019,
  author={Denaro, Lino Garda and Hung Lin, Chao},
  booktitle={IGARSS 2019 - 2019 IEEE International Geoscience and Remote Sensing Symposium}, 
  title={{N}onlinear Relative Radiometric Normalization For {L}andsat-7 and {L}andsat-8 Imagery}, 
  year={2019},
  volume={},
  number={},
  pages={1967-1969},
  doi={10.1109/IGARSS.2019.8900651}}

@article{Denaro2018,
  author = {Denaro, Lino Garda and Lin, Bo-Yi and Syariz, Muhammad Aldila and Jaelani, Lalu Muhamad and Lin, Chao-Hung},
  title = {Pseudoinvariant feature selection for cross-sensor optical satellite images},
  journal = {Journal of Applied Remote Sensing},
  volume = {12},
  number = {4},
  pages = {045002},
  year = {2018},
  month = {Oct.},
  doi = {10.1117/1.JRS.12.045002},
  url = {https://doi.org/10.1117/1.JRS.12.045002}
}

@InProceedings{pmlrandrew13,
  title = 	 {Deep Canonical Correlation Analysis},
  author = 	 {Andrew, Galen and Arora, Raman and Bilmes, Jeff and Livescu, Karen},
  booktitle = 	 {Proc. International Conference on Machine Learning},
  pages = 	 {1247--1255},
  year = 	 {2013},
  editor = 	 {Dasgupta, Sanjoy and McAllester, David},
  volume = 	 {28},
  number =       {3},
  series = 	 {Proceedings of Machine Learning Research},
  address = 	 {Atlanta, Georgia, USA},
  month = 	 {17--19 Jun},
  publisher =    {PMLR},
  url = 	 {https://proceedings.mlr.press/v28/andrew13.html}
}

@inproceedings{benton2019deep,
    title = "Deep Generalized Canonical Correlation Analysis",
    author = "Benton, Adrian  and
      Khayrallah, Huda  and
      Gujral, Biman  and
      Reisinger, Dee Ann  and
      Zhang, Sheng  and
      Arora, Raman",
    editor = "Augenstein, Isabelle  and
      Gella, Spandana  and
      Ruder, Sebastian  and
      Kann, Katharina  and
      Can, Burcu  and
      Welbl, Johannes  and
      Conneau, Alexis  and
      Ren, Xiang  and
      Rei, Marek",
    booktitle = "Proceedings of the 4th Workshop on Representation Learning for NLP (RepL4NLP-2019)",
    month = aug,
    year = "2019",
    address = "Florence, Italy",
    publisher = "Association for Computational Linguistics",
    url = "https://aclanthology.org/W19-4301/",
    doi = "10.18653/v1/W19-4301",
    pages = "1--6"
}

@InProceedings{pmlrwangb15,
  title = 	 {On Deep Multi-View Representation Learning},
  author = 	 {Wang, Weiran and Arora, Raman and Livescu, Karen and Bilmes, Jeff},
  booktitle = 	 {Proc. International Conference on Machine Learning},
  pages = 	 {1083--1092},
  year = 	 {2015},
  editor = 	 {Bach, Francis and Blei, David},
  volume = 	 {37},
  series = 	 {Proceedings of Machine Learning Research},
  address = 	 {Lille, France},
  month = 	 {07--09 Jul},
  publisher =    {PMLR},
  url = 	 {https://proceedings.mlr.press/v37/wangb15.html}
}

@INPROCEEDINGS{Chen2021,
  author={Chen, Xinlei and He, Kaiming},
  booktitle={2021 IEEE/CVF Conference on Computer Vision and Pattern Recognition (CVPR)}, 
  title={Exploring Simple Siamese Representation Learning}, 
  address ={Nashville, TN, USA},
  year={20-25 June 2021},
  volume={},
  number={},
  pages={15745-15753},
  doi={10.1109/CVPR46437.2021.01549}}

@inproceedings{NEURIPS2020_f3ada80d,
 author = {Grill, Jean-Bastien and Strub, Florian and Altch\'{e}, Florent and Tallec, Corentin and Richemond, Pierre and Buchatskaya, Elena and Doersch, Carl and Avila Pires, Bernardo and Guo, Zhaohan and Gheshlaghi Azar, Mohammad and Piot, Bilal and kavukcuoglu, koray and Munos, Remi and Valko, Michal},
 booktitle = {Advances in Neural Information Processing Systems},
 editor = {H. Larochelle and M. Ranzato and R. Hadsell and M.F. Balcan and H. Lin},
 pages = {21271--21284},
 publisher = {Curran Associates, Inc.},
 title = {Bootstrap Your Own Latent - A New Approach to Self-Supervised Learning},
 url = {https://proceedings.neurips.cc/paper_files/paper/2020/file/f3ada80d5c4ee70142b17b8192b2958e-Paper.pdf},
 volume = {33},
 year = {2020}
}

@InProceedings{pmlrzbontar21a,
  title = 	 {Barlow Twins: Self-Supervised Learning via Redundancy Reduction},
  author =       {Zbontar, Jure and Jing, Li and Misra, Ishan and LeCun, Yann and Deny, Stephane},
  booktitle = 	 {Proc. International Conference on Machine Learning},
  pages = 	 {12310--12320},
  year = 	 {2021},
  editor = 	 {Meila, Marina and Zhang, Tong},
  volume = 	 {139},
  series = 	 {Proceedings of Machine Learning Research},
  month = 	 {18--24 Jul},
  publisher =    {PMLR},
  url = 	 {https://proceedings.mlr.press/v139/zbontar21a.html}
}

@article{liu2025cd,
  title={{CD-STMamba: T}owards remote sensing image change detection with spatio-temporal interaction mamba model},
  author={Liu, Shanwei and others},
  journal={IEEE J. Sel. Topics Appl. Earth Observ. in Remote Sens.},
  year={Apr. 2025},
  pages={10471--10485},
  volume={18},
  publisher={IEEE}
}

@article{DU2002123,
title = {Radiometric normalization of multitemporal high-resolution satellite images with quality control for land cover change detection},
journal = {Remote Sensing of Environment},
volume = {82},
number = {1},
pages = {123-134},
year = {Sep. 2002},
issn = {0034-4257},
doi = {https://doi.org/10.1016/S0034-4257(02)00029-9},
url = {https://www.sciencedirect.com/science/article/pii/S0034425702000299},
author = {Yong Du and Philippe M Teillet and Josef Cihlar}
}

@ARTICLE{CODEHCD,
  author={Lin, Tzu-Hsuan and Lin, Chia-Hsiang},
  journal={IEEE Trans. Geosci. Remote Sens.}, 
  title={Hyperspectral Change Detection Using Semi-Supervised Graph Neural Network and Convex Deep Learning}, 
  year={Jun. 2023},
  volume={61},
  number={},
  pages={1--18},
}

@article{PAINT,
  title={Interpretable Landsat-to-Hyperspectral Dual Super-Resolution Without Large Matrix Inversion},
  author={Lin, Chia-Hsiang and Huang, Jian-Kai and Young, Si-Sheng and Zheng, Wei-Cheng},
  journal={IEEE J. Sel. Topics Appl. Earth Observ. Remote Sens. Observations and Remote Sensing},
  year={Aug. 2026},
  pages={1--16},
  volume={},
  publisher={IEEE}
}

@ARTICLE{Bovolo2007polar,
  author={Bovolo, Francesca and Bruzzone, Lorenzo},
  journal={IEEE Trans. Geosci. Remote Sens.}, 
  title={A Theoretical Framework for Unsupervised Change Detection Based on Change Vector Analysis in the Polar Domain}, 
  year={Jan. 2007},
  volume={45},
  number={1},
  pages={218-236},
  doi={10.1109/TGRS.2006.885408}}

@ARTICLE{Wu2014sfa,
  author={Wu, Chen and Du, Bo and Zhang, Liangpei},
  journal={IEEE Trans. Geosci. Remote Sens.}, 
  title={Slow Feature Analysis for Change Detection in Multispectral Imagery}, 
  year={Jul. 2014},
  volume={52},
  number={5},
  pages={2858-2874},
  doi={10.1109/TGRS.2013.2266673}}

@ARTICLE{Gao2006blending,
  author={Feng Gao and Masek, J. and Schwaller, M. and Hall, F.},
  journal={IEEE Trans. Geosci. Remote Sens.}, 
  title={On the blending of the Landsat and MODIS surface reflectance: predicting daily Landsat surface reflectance}, 
  year={Aug. 2006},
  volume={44},
  number={8},
  pages={2207-2218},
  doi={10.1109/TGRS.2006.872081}}

@ARTICLE{Latifovic2003brdf,
  author={Latifovic, R. and Cihlar, J. and Jing Chen},
  journal={IEEE Trans. Geosci. Remote Sens.}, 
  title={A comparison of BRDF models for the normalization of satellite optical data to a standard Sun-target-sensor geometry}, 
  year={Aug. 2003},
  volume={41},
  number={8},
  pages={1889-1898},
  doi={10.1109/TGRS.2003.811557}}

@ARTICLE{Vermote1997sixs,
  author={Vermote, E.F. and Tanre, D. and Deuze, J.L. and Herman, M. and Morcette, J.-J.},
  journal={IEEE Trans. Geosci. Remote Sens.}, 
  title={Second Simulation of the Satellite Signal in the Solar Spectrum, 6{S}: {A}n overview}, 
  year={May 1997},
  volume={35},
  number={3},
  pages={675-686},
  doi={10.1109/36.581987}}

@ARTICLE{Markham2004landsat,
  author={Markham, B.L. and Storey, J.C. and Williams, D.L. and Irons, J.R.},
  journal={IEEE Trans. Geosci. Remote Sens.}, 
  title={Landsat sensor performance: {H}istory and current status}, 
  year={Dec. 2004},
  volume={42},
  number={12},
  pages={2691-2694},
  doi={10.1109/TGRS.2004.840720}}

@ARTICLE{HyperCSI,
  author={Lin, Chia-Hsiang and Chi, Chong-Yung and Wang, Yu-Hsiang and Chan, Tsung-Han},
  journal={IEEE Trans. Signal Process.}, 
  title={A fast hyperplane-based minimum-volume enclosing simplex algorithm for blind hyperspectral unmixing}, 
  year={Dec. 2015},
  volume={64},
  number={8},
  pages={1946--1961},
  doi={10.1109/TSP.2015.2508778}}

@article{CODE,
  title={{ADMM-ADAM}: {A} new inverse imaging framework blending the advantages of convex optimization and deep learning},
  author={Lin, Chia-Hsiang and Lin, Yen-Cheng and Tang, Po-Wei},
  journal={IEEE Trans. Geosci. Remote Sens.},
  volume={60},
  pages={1--16},
  year={Sep. 2021},
  publisher={IEEE}
}

@ARTICLE{CODEMM,
  author={Lin, Chia-Hsiang and Chu, Man-Chun and Tang, Po-Wei},
  journal={IEEE Trans. Geosci. Remote Sens.}, 
  title={{CODE-MM}: {C}onvex Deep Mangrove Mapping Algorithm Based on Optical Satellite Images}, 
  year={Sep. 2023},
  volume={61},
  pages={1--19},
  doi={10.1109/TGRS.2023.3314088}}

@article{lin2015identifiability,
  title={Identifiability of the simplex volume minimization criterion for blind hyperspectral unmixing: The no-pure-pixel case},
  author={Lin, Chia-Hsiang and Ma, Wing-Kin and Li, Wei-Chiang and Chi, Chong-Yung and Ambikapathi, ArulMurugan},
  journal={IEEE Trans. Geosci. Remote Sens.},
  volume={53},
  number={10},
  pages={5530--5546},
  year={Oct. 2015},
  publisher={IEEE}
}

@ARTICLE{Xie2022global,
  author={Xie, Shuai and Sun, Lin and Liu, Liangyun and Liu, Xiaomi},
  journal={IEEE Trans. Geosci. Remote Sens.}, 
  title={Global Cross-Sensor Transformation Functions for {L}andsat-8 and {S}entinel-2 Top of Atmosphere and Surface Reflectance Products Within Google Earth Engine}, 
  year={Apr. 2022},
  volume={60},
  number={},
  pages={1-9},
  doi={10.1109/TGRS.2022.3170338}}

@ARTICLE{Liu2015,
  author={Liu, Sicong and Bruzzone, Lorenzo and Bovolo, Francesca and Du, Peijun},
  journal={IEEE Trans. Geosci. Remote Sens.}, 
  title={Hierarchical Unsupervised Change Detection in Multitemporal Hyperspectral Images}, 
  year={Jan. 2015},
  volume={53},
  number={1},
  pages={244-260},
  doi={10.1109/TGRS.2014.2321277}}

@article{wang2025harmu,
  title={HARMU: {A} multiband sensor harmonization for building virtual constellations. Application to {L}andsat-8 and {S}entinel-2},
  author={Wang, Changjing and Yin, Gaofei and Fu, Rui and Descals, Adri{\`a} and Li, Wenjuan and Weiss, Marie and Baret, Fr{\'e}d{\'e}ric and Verger, Aleixandre},
  journal={IEEE Trans. Geosci. Remote Sens.},
  year={Jun. 2025},
  publisher={IEEE}
}

@ARTICLE{Shen2024spectral,
  author={Shen, Aojie and Shao, Jie},
  journal={IEEE Trans. Geosci. Remote Sens.}, 
  title={Spectral Harmonization {L}andsat-8 and {S}entinel-2{A}: {T}he Matching Bands Adjustment Before the Missing Bands Prediction Method}, 
  year={Jan. 2024},
  volume={62},
  number={},
  pages={1-14},
  doi={10.1109/TGRS.2024.3354239}}

@ARTICLE{Alparone2025harmonization,
  author={Alparone, Luciano and Arienzo, Alberto and Garzelli, Andrea},
  journal={IEEE Trans. Geosci. Remote Sens.}, 
  title={Harmonization of Multispectral and Panchromatic Images From {VHR}/{EHR} Spaceborne Scanners}, 
  year={Sep. 2025},
  volume={63},
  number={},
  pages={1-16},
  doi={10.1109/TGRS.2025.3605739}}

@ARTICLE{Zhao2026terrain-aware,
  author={Zhao, Yujie and Jia, Guorui and Zhao, Huijie},
  journal={IEEE Trans. Geosci. Remote Sens.},
  title={{T}errain-aware Uncertainty Quantification and Cross-Sensor Consistency Analysis of Hyperspectral Surface Reflectance},
  year={Dec. 2026},
  volume={64},
  pages={1--18},
  doi={10.1109/tgrs.2025.3649800}}

@ARTICLE{Lu2025mapping,
  author={Lu, Wei and Hu, Yunfeng and {Batunacun} and Liu, Jia and Li, Hao},
  journal={IEEE Trans. Geosci. Remote Sens.},
  title={{M}apping of Hay-Harvesting Grasslands Using Harmonized {L}andsat {S}entinel-2 Time Series and Deep Learning in Temperate Steppe},
  year={Sep. 2025},
  volume={63},
  pages={1--16},
  doi={10.1109/tgrs.2025.3614355}}

@ARTICLE{Saha2019unsupervised,
  author={Saha, Sudipan and Bovolo, Francesca and Bruzzone, Lorenzo},
  journal={IEEE Trans. Geosci. Remote Sens.},
  title={{U}nsupervised Deep Change Vector Analysis for Multiple-Change Detection in {VHR} Images},
  year={Jan. 2019},
  volume={57},
  pages={3677--3693},
  doi={10.1109/tgrs.2018.2886643}}

@ARTICLE{Chen2022unsupervised,
  author={Chen, Hongruixuan and Yokoya, Naoto and Wu, Chen and Du, Bo},
  journal={IEEE Trans. Geosci. Remote Sens.},
  title={{U}nsupervised Multimodal Change Detection Based on Structural Relationship Graph Representation Learning},
  year={Dec. 2022},
  volume={60},
  pages={1--18},
  doi={10.1109/tgrs.2022.3229027}}

@ARTICLE{Zhang2023self-supervised,
  author={Zhang, Yuxiang and Zhao, Yang and Dong, Yanni and Du, Bo},
  journal={IEEE Trans. Geosci. Remote Sens.},
  title={{S}elf-Supervised Pretraining via Multimodality Images With Transformer for Change Detection},
  year={Apr. 2023},
  volume={61},
  pages={1--11},
  doi={10.1109/tgrs.2023.3271024}}

@ARTICLE{Jian2024uncertainty-aware,
  author={Jian, Ping and Ou, Yimin and Chen, Keming},
  journal={IEEE Trans. Geosci. Remote Sens.},
  title={{U}ncertainty-Aware Graph Self-Supervised Learning for Hyperspectral Image Change Detection},
  year={Feb. 2024},
  volume={62},
  pages={1--19},
  doi={10.1109/tgrs.2024.3363886}}

@ARTICLE{Li2025unsupervised,
  author={Li, Weisong and Li, Yinwei and Zhu, Yiming and Wang, Haipeng},
  journal={IEEE Trans. Geosci. Remote Sens.},
  title={{U}nsupervised {SAR} Image Change Detection via Structure Feature-Based Self-Representation Learning},
  year={Aug. 2025},
  volume={63},
  pages={1--18},
  doi={10.1109/tgrs.2025.3598588}}

\begin{IEEEbiography}[{\resizebox{1in}{!}{\includegraphics[width=1in,height=1.25in,clip,keepaspectratio]{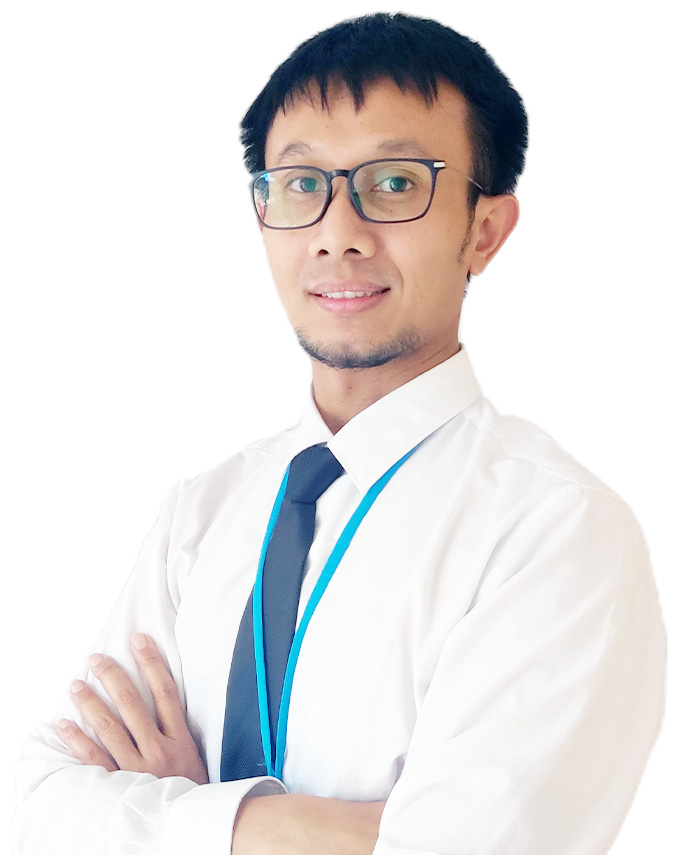}}}]
{\bf Lino Garda Denaro}
(S'20)
received the B.Eng. (S.T.) degree in geomatics engineering from the Institute of Technology Sepuluh Nopember (ITS), Surabaya, Indonesia, in 2015, and the M.S. and Ph.D. degrees from the Department of Geomatics, National Cheng Kung University (NCKU), Tainan, Taiwan, in 2018 and 2022, respectively.

From 2022 to 2025, he was a Postdoctoral Fellow at National Taiwan University (NTU), Taipei, Taiwan. In 2026, he joined the Department of Electrical Engineering, National Cheng Kung University (NCKU), Tainan, Taiwan, as a Research Fellow (visiting research program). He is currently a Researcher at the Research Center for Geoinformatics, National Research and Innovation Agency (BRIN), Indonesia. He is also a member of the BRIN Quick Response Task Force, contributing to national post-disaster strategic planning and natural hazard mitigation frameworks in Indonesia.
His research interests include image processing, optimization, computer vision, machine learning, artificial intelligence (transformers, graph neural network, quantum-based artificial intelligences), remote sensing, and geoinformatics-related applications. 

\end{IEEEbiography}

\begin{IEEEbiography}[{\resizebox{0.9in}{!}{\includegraphics[width=1in,height=1.25in,clip,keepaspectratio]{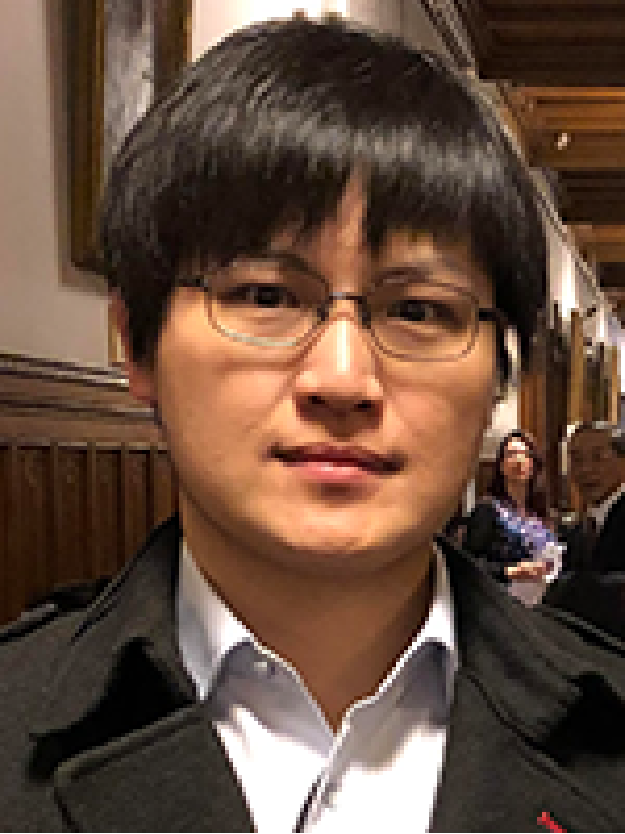}}}]
{\bf Chia-Hsiang Lin}
(S'10-M'18-SM'24)
received the B.S. degree in electrical engineering and the Ph.D. degree in communications engineering from National Tsing Hua University (NTHU), Taiwan, in 2010 and 2016, respectively.
From 2015 to 2016, he was a Visiting Student of Virginia Tech,
Arlington, VA, USA.

He is currently a Professor with the Department of Electrical Engineering,
National Cheng Kung University (NCKU), Taiwan, and also serves as a Technical Director of Smart Sensing \& Systems Technology Center, Industrial Technology Research Institute (ITRI).
Before joining NCKU, he held research positions with The Chinese University of Hong Kong, HK (2014 and 2017),
NTHU (2016-2017),
and the University of Lisbon (ULisboa), Lisbon, Portugal (2017-2018).
He was an Assistant Professor with the Center for Space and Remote Sensing Research, National Central University, Taiwan, in 2018, a Visiting Professor with ULisboa, in 2019, and a Visiting Professor with Texas A\&M University, USA, in 2025.
His research interests include network science,
quantum computing,
convex geometry and optimization, blind signal processing, and imaging science.

Dr. Lin received the Emerging Young Scholar Award (The 2030 Cross-Generation Program) from National Science and Technology Council (NSTC), from 2023 to 2027,
the Future Technology Award from NSTC, in 2022 and 2026,
the Outstanding Youth Electrical Engineer Award from The Chinese Institute of Electrical Engineering (CIEE), in 2022,
the Best Young Professional Member Award from IEEE Tainan Section, in 2021,
and the Prize Paper Award from IEEE Geoscience and Remote Sensing Society (GRS-S), in 2020.
He received the Ministry of Science and Technology (MOST) Young Scholar Fellowship, together with the EINSTEIN Grant Award, from 2018 to 2023.
In 2016, he was a recipient of the Outstanding Doctoral Dissertation Award from the Chinese Image Processing and Pattern Recognition Society and the Best Doctoral Dissertation Award from the IEEE GRS-S.
\end{IEEEbiography}

\end{document}